\documentclass[onecolumn,journal]{IEEEtran}

\usepackage{setspace}
\def\BibTeX{{\rm B\kern-.05em{\sc i\kern-.025em b}\kern-.08em T\kern-.1667em\lower.7ex\hbox{E}\kern-.125emX}}

\ifCLASSINFOpdf
\else
\fi
\usepackage{graphicx}
\usepackage{latexsym}
\usepackage{hyperref}
\usepackage{epstopdf}
\usepackage{multicol}
\usepackage{amsmath}
\usepackage{mathtools}
\usepackage{amssymb}
\usepackage{esint}
\usepackage{amsfonts}
\usepackage{multirow}
\usepackage{wrapfig}
\usepackage{cite}
\usepackage{subcaption}
\usepackage{lipsum}
\usepackage{float}    
\usepackage{stfloats} 
\usepackage{algorithm}
\usepackage[noend]{algpseudocode}
\usepackage{caption}
\usepackage[none]{hyphenat}
\usepackage[utf8]{inputenc}
\usepackage[T1]{fontenc}
\usepackage[english]{babel} 
\usepackage{amsmath,amsfonts,amsthm,bm} 
\usepackage{ifthen}

\usepackage{epsfig}
\usepackage{psfrag}
\usepackage{nomencl}

\newtheorem{prop}{Proposition}
\usepackage{caption}
\usepackage{subcaption}

\usepackage{xcolor}

\begin{document}

\title{Terahertz Multiple Access: A Deep Reinforcement Learning Controlled Multihop IRS Topology}

\author{Muhammad Shehab,~\IEEEmembership{Member,~IEEE,}
        Abdullateef Almohamad,~\IEEEmembership {Member,~IEEE,}
        Mohamed Elsayed,~\IEEEmembership{Member,~IEEE,}
        Ahmed Badawy, ~\IEEEmembership{Member,~IEEE,} 
        Tamer Khattab,~\IEEEmembership{Senior Member,~IEEE,}
        Nizar Zorba,~\IEEEmembership{Senior Member,~IEEE}
        Mazen Hasna,~\IEEEmembership{Senior Member,~IEEE},
        Daniele Trinchero.

\thanks{M. Shehab, M. Elsayed, T. Khattab, N. Zorba, and M. Hasna are with the Electrical Engineering, Qatar University, Doha, Qatar. A. Badawy is with the Computer Science and Engineering, Qatar University, Doha, Qatar. A. Almohamad is with the Electrical and Computer Engineering, Texas A\&M University Qatar, Doha, Qatar. D. Trinchero is with Dipartimento di Elettronica, Politecnico di Torino, Torino, Italy, e-mail: MuhammadShehab@ieee.org, abdullateef@ieee.org, hamid@qu.edu.qa, badawy@qu.edu.qa, tkhattab@ieee.org, nizarz@qu.edu.qa, hasna@qu.edu.qa, and daniele.trinchero@polito.it. This research work was made possible by grant number AICC03-0530-200033 from the Qatar National Research Fund (QNRF).  Statements made herein are the sole responsibility of the authors.}} 

\maketitle

\IEEEpeerreviewmaketitle

\begin{abstract}
We investigate THz communication uplink multiple access using cascaded intelligent reflecting surfaces (IRSs) assuming correlated channels. Two independent objectives to be achieved via adjusting the phases of the cascaded IRSs: 1) maximizing the received rate of a desired user under interference from the second user and 2) maximizing the sum rate of both users. The resulting optimization problems are non-convex. For the first objective, we devise a sub-optimal analytical solution by maximizing the received power of the desired user, however, this results in an over determined system.  Approximate solutions using pseudo-inverse and block-based approaches are attempted. For the second objective, a loose upperbound is derived and an exhaustive search solution is utilized. We then use deep reinforcement learning (DRL) to solve both objectives. Results reveal the suitability of DRL for such complex configurations. For the first objective, the DRL-based solution is superior to the sub-optimal mathematical methods, while for the second objective, it produces sum rates almost close to the exhaustive search. Further, the results reveal that as the correlation-coefficient increases, the sum rate of DRL increases, since it benefits from the presence of correlation in the channel to improve statistical learning. 
\end{abstract}

 
\makenomenclature
\mbox{}
\nomenclature{THz}{Terahertz}
\nomenclature{IRS}{Intelligent Reflecting Surfaces}
\nomenclature{BS}{Base Station}
\nomenclature{MIMO}{Multiple-Input and Multiple-Output}
\nomenclature{LoS}{line-of-sight}
\nomenclature{DF}{Decode-and-forward}
\nomenclature{DRL}{Deep Reinforcement Learning}
\nomenclature{DDPG}{Deep Deterministic Policy Gradient}
\nomenclature{AWGN}{Additive white Gaussian noise}
\nomenclature{RU}{Reflecting Unit}
\nomenclature{FNBW}{First Null Beamwidth}
\nomenclature{FSPL}{Free Space Path Loss}
\nomenclature{Tx}{Transmitter}
\nomenclature{Rx}{Receiver}
\printnomenclature [6em]

\section{Introduction}

\IEEEPARstart{T}{he} sixth generation (6G) wireless communications need to provide radically modern services and bandwidth-intensive applications compared to the fifth-generation (5G) such as immersive remote presence,
connected robotics; autonomous systems (CRAS), immersive extended reality (XR), and digital twin. These applications demand a 1000$\times$ increase in capacity compared to 5G mobile systems \cite{IEEEhowto:1}. To achieve these requirements and overcome the conflict between service demands and spectrum scarcity \cite{IEEEhowto:2}, there is a need to boost the current wireless spectrum bands, and migrate towards higher terahertz (THz) frequency bands which range from 0.1 THz to 10 THz. These frequencies are considered a key element in 6G wireless communications because they possibly support considerable capacities and data rates. However, high-frequency values lead to severe path attenuation, high propagation losses, and sporadic wireless links. Further, these values produce very small wavelength ($\lambda$) values, which in turn results in very short communication distances and increases the susceptibility to molecular absorption and blockage \cite{IEEEhowto:1}. To improve the received signal power, and achievable data rate, this paper investigates the intelligent reflecting surface (IRS) as an emerging technology and promising solution \cite{IEEEhowto:3}. IRS manipulates the incident electromagnetic waves and adjusts the phase shifts of the semi-passive reflecting elements in a programmable behavior to yield a smart radio environment and enhance the data rate in an energy-efficient and cost-effective manner \cite{IEEEhowto:4}.

Many recent studies examined IRS deployment in THz communications to inspect the power of IRS to improve the coverage and achievable data rate \cite{IEEEhowto:5} - \cite{IEEEhowto:11}. For instance, the authors in \cite{IEEEhowto:5} studied the scenario of IRS-assisted multi-hop multi-pair unicast network, where multiple sources are communicating with multiple destinations. They proposed distributed multiple IRSs controls for a multi-hop interference channel with the purpose of maximizing the achievable rate. A multi-IRS assisted massive multiple-input multiple-output system was studied \cite{IEEEhowto:6} to increase the minimum received signal power, where a base station (BS) equipped with multi-antenna transmits independent signals to a group of remote users equipped with single-antenna, and a cascaded line-of-sight (LOS) communication links are established among the BS and users by using the cooperative signal reflections of various IRSs groups. In \cite{IEEEhowto:7} the authors examined the performance of the rate achieved for a decode-and-forward (DF) relaying assisted multi-IRS system, where a single source is communicating with a single destination (user) with the objective of obtaining the optimal IRSs configuration, number of IRSs, and number of IRS reflecting elements that maximize the ergodic rate. Further, in \cite{IEEEhowto:8} the authors considered a multi-hop IRS-assisted multi-user downlink communication scenario, where the BS is communicating to $K$ users with the scope of maximizing the sum rate by jointly optimizing the beamforming at the BS, and the multiple IRS phase shift reflection matrices. Moreover, in \cite{IEEEhowto:9}, the authors investigated the scenario of an uplink multi-hop IRS communication system where multi-users (sources) are communicating with a single destination. Their scope was to extend the link range in THz communications and maximize the power at the receiver. They proposed a cascaded passive IRS THz system to overcome the high propagation losses caused by the absorption in the air molecules.   

To this end, the aforementioned research papers \cite{IEEEhowto:5} - \cite{IEEEhowto:9} adopted mathematical techniques to solve their optimization problem. Unlike these studies, the research papers in \cite{IEEEhowto:10} and \cite{IEEEhowto:11} utilized the DRL algorithm to solve the non-convex optimization problem. The authors proposed a hybrid beamforming scheme for the cascaded IRS-aided networks to enhance the coverage of the THz communication links. They investigated the joint design of the analog beamforming at the IRSs and the digital beamforming at the BS to overcome the propagation loss in THz downlink broadcast system, which is a single source to a multi-destination (multi-user) scenario.

\subsection{Contributions}
To extend the range of the THz links and compensate for the losses at such a high operating frequency range, we adopt multi-hop IRSs (also referred to as cascaded IRSs) as the core component in our system model. Since, this is typical environment for users using THz links where the coverage is small, then we only consider small areas where several users are not expected. Thus, we formulate an optimization problem under the assumption of a two-user system, where the objective is to find the optimum phase shifts of the multi-hop IRS's element in order to maximize the received rate for any specific user, and the sum rate for both users. The major challenge in our scheme lies in the non-convexity of the objective function due to the constant modulus constraints of the reflecting IRS elements, non-linear constraints, and computationally intractable multi-hop links. In general, the optimal solution to this NP-hard problem is unknown, and it is difficult to derive an analytical solution using traditional mathematical methods, and the exhaustive search is not practical for large-scale communication systems \cite{IEEEhowto:10}, \cite{IEEEhowto:11}. In addition, solving the problem of the received rate maximization leads to an over-determined system. To tackle this, we exploit a Deep Deterministic Policy Gradient (DDPG)-based algorithm, which is a DRL method, to obtain feasible solutions.

To the best of our knowledge, none of the existing research papers in the literature leveraged the DRL method to solve the over-determined system of equations for the uplink cascaded IRS multiple access scenario. In this paper, we address the above-mentioned gap in the literature by employing DRL- in particular, DDPG, to jointly optimize the phase shifts of each IRS in the cascaded IRS system taking into consideration the case of the spatially correlated channel \cite{IEEEhowto:12} between IRS$_1$ and IRS$_2$, to achieve two main objectives: a) maximizing the rate for any specific user, and b) maximizing the sum rate for both users. 

We detail our contributions below against our two main objectives under the consideration of multi-hop IRSs and multiple access systems operating in the THz range:

\begin{itemize}

\item Our first objective is to maximize the rate for any specific user under the assumption that the second user is considered as interference. For this objective:
\begin{itemize}
\item We formulate the cascaded IRS phase shift optimization problem that includes IRS$_1$ and IRS$_2$ phases as a closed form, and we prove that it is non-convex and mathematically intractable.
\item Further, to overcome this, we propose two sub-optimal solutions, where our objective is to find the optimal phases of the IRS$_1$ and IRS$_2$ elements that maximize the received power of the desired user. We show that when solving for the optimized phase shifts that maximize the received power of the desired user, the system is over-determined and therefore, we propose two solutions for this problem through the use of pseudo-inverse and block solutions.
\item Moreover, we design a DDPG algorithm to find the optimum phases that maximize the rate of the desired user.
\end{itemize}

\item Our second objective is to maximize the sum rate of two users, For this objective:
\begin{itemize}
\item We provide analytical analysis for the problem in the case of cascaded IRSs.
\item We design a DDPG algorithm to obtain the optimum phase shifts of the cascaded IRSs that maximize the received sum rate.
\end{itemize}
\item We simulate our proposed solutions and compare the obtained results to those obtained through the exhaustive search algorithm and through randomly generated phase shifts.
\end{itemize}





\subsection{Paper organization}

The remainder of this research paper is structured as follows; section II describes the system and channel model for the multi-hop IRS scenario. Section III presents the problem formulation for maximizing the rate of the desired user operating under interference, and Section IV discusses sum rate maximization problem for both users. Thus, it discusses the end-to-end sum rate derivation in the cascaded IRS scenario. Section V introduces the proposed solution using DDPG for the cascaded IRS phase control. The simulation results are discussed in section VI, and section VII concludes the paper. \\

Notation: For more convenience, frequent symbols and parameters along with their description are illustrated in Table \ref{table:1}.

\renewcommand\thetable{1}
\begin{table*} 
\caption{List of frequently used parameters and symbols.} 
\centering 
\begin{tabular}{|c c|c c|} 
\hline
& \textbf{Parameters and Symbols} &  \textbf{Description} &\\ 
\hline
& $x_k$ & Transmitted signal for each user $k$ &\\ 
\hline
& $y_k$ & Received signal for each user $k$ &\\ 
\hline
& $z_k$ & Signal for each user $k$ &\\ 
\hline
& $P_t$ & Transmit power for each user &\\ 
\hline
& $\lambda$ &  Wavelength &\\ 
\hline
&$\mathbf{h}_{r}$ & Channel between IRS2 and the receiver &\\   
\hline
&$\mathbf{H}_{m,n}$ & Channel between IRS$_1$ and IRS$_2$  &\\ 
\hline
&$\mathbf{h}_{t,k}$ & Channel between each user $k$ and IRS$_1$&\\ 
\hline
& $n_0$ &  AWGN in linear scale &\\ 
\hline
& $K_1$, $K_2$ &  Rician Factor for the transmitter channel and receiver channel&\\ 
\hline
& $\mathbf{R}$  &  Covariance Matrix &\\ 
\hline
& $\mathbf{\Phi}_M$, $\mathbf{\Phi}_N$ &  Phase shift reflection matrix for IRS$_1$ and IRS$_2$, respectively &\\ 
\hline
& $\phi_m$, $\phi_n$ & Phase shift of IRS$_1$ reflecting element $m$ and IRS$_2$ reflecting element $n$, respectively &\\
\hline
& $\eta_m$  &  Phase shift of $m^{th}$ IRS$_1$ reflecting element &\\ 
\hline
& $\psi_n$  &  Phase shift of $n^{th}$ IRS$_2$ reflecting element &\\ 
\hline
& $D_t$, $D_r$  &  Antenna diameters for each $T_{x_k}$ and $R_x$, respectively &\\ 
\hline
& $r_{tk}$, $r_2$, $r_3$  &  Distance between: each user $k$ and IRS$_1$, IRS$_1$ and IRS$_2$, and IRS$_2$ and $R_x$, respectively &\\ 
\hline
& $r_{k,1,h}$, $r_{2,h}$, $r_{3,h}$ &  Horizontal distance between: user $k$ and center of IRS$_1$, centers of IRS$_1$ and IRS$_2$,&\\ && and center of IRS$_2$ and $R_x$, respectively &\\
\hline
& $\theta_{ik,1}$ & Incident angle from user $k$ w.r.t. the center of the illuminated area at IRS$_1$
&\\ 
\hline
& $\theta_{r,1}$  & Reflected angle w.r.t. the center of the illuminated area at IRS$_1$  &\\ 
\hline
& $\theta_{i,2}$  & Incident angle from IRS$_1$ w.r.t. the center of the illuminated area at IRS$_2$  &\\ 
\hline
& $\theta_{r,2}$  & Reflected angle w.r.t. the center of the illuminated area at IRS$_2$ &\\ 
\hline 
& $\ell_{Tx,k}$, $\ell_{Rx}$, $\ell_{s1}$, $\ell_{s2}$ & The height of the $T_{x_k}$, $R_x$, IRS$_1$ and IRS$_2$, respectively&\\ 
\hline
& M, N  & Number of reflecting elements for IRS$_1$ and IRS$_2$ respectiveky &\\ 
\hline
& $[\mathbf{R}]_{m,n}$  & Covariance matrix obtained based on the exponential spatial correlation model &\\ 
\hline
& $\rho^{|m-n|}$  & Correlation-coefficient among the adjacent RUs &\\ 
\hline
& $\Omega_k$ & Phase shifts corresponding to signal traveled from user $k$ to IRS$_1$ &\\ 
\hline
& $\Omega_3$  & Phase shift corresponding to signal traveled from IRS2 to $R_x$ &\\ 
\hline
& $o$  & Angle measured from the broadside of the antenna  &\\ 
\hline
& $G_t(o)$, $G_r(o)$ & Gains for the users' and receiver antennas respectively &\\ 
\hline
& $e_t$, $e_r$ & Aperture Efficiencies for the $T_{x_k}$ and $R_x$ &\\ 
\hline
 &  $L_{\tau,k}$ & Total losses and gains on the path between each $T_{x_k}$ and the $R_x$ &\\ 
\hline
 &  $L_{FSPL,\tau,k}$ & Total FSPL for $T_{x_k}$ &\\ 
\hline
 &  $L_{abs,\tau,k}$ & Total absorption loss for $T_{x_k}$ &\\ 
\hline
 &  $G_{t,k}$ $G_{r}$ & $T_{x_k}$ , and $R_x$ antenna gains &\\ 
\hline
 &  $G(\theta_{i1,k})$, $(\theta_{r,1})$ & Gain of IRS$_1$ RU from the incident and reflection angles &\\ 
\hline 
&$P_{r,m}$ & The power reflected from the $m_{th}$ RU of IRS$_1$ &\\ 
\hline 
&$P_{r,mn}$ & The power reflected from the $n_{\textrm{th}}$ RU of IRS2 because of being &\\ && illuminated by the signal reflected by the $m_{\textrm{th}}$ RU of IRS$_1$ &\\ 
\hline 
&$P_{rx,mn}$ & The received captured power at the $R_x$ &\\ 
\hline 
&$P_{Rx_k}$&  The total received power for user $k$ at the receiver ($R_x$ ) &\\ 
\hline
& $\alpha_m$ & The reflection coefficient of the $m^{th}$ RU of IRS$_1$ &\\ 
\hline 
& $\alpha_n$ & The reflection coefficient of the $n^{th}$ RU of IRS$_2$ &\\ 
\hline
& $\varphi_{t_{k_m}}$  & The phase for the transmitter channel for user $k$ and $m^{th}$ RU &\\ 
\hline 
& $\varphi_{mn}$ & The phase for the $\mathbf{h}_{m,n}$ channel for $m^{th}$ and $n^{th}$ RU &\\ 
\hline 
& $\varphi_{r_n}$ & The phase for the receiver channel for $ n^{th} RU$ &\\ 
\hline 
& $\large \gamma_k$ & The received SINR for the $T_{x_k}$ at the $R_x$  &\\ 
\hline  
&$R_k$,  & The data rate for user $k$  &\\
\hline  
&$R_{sum}$ & and in practical and in practical &\\
\hline  
&$\Xi$ & Phase search steps &\\
\hline  
\end{tabular}
\label{table:1} 
\end{table*} 

\begin{figure*}
    \centering
    \includegraphics[width=0.9\linewidth]{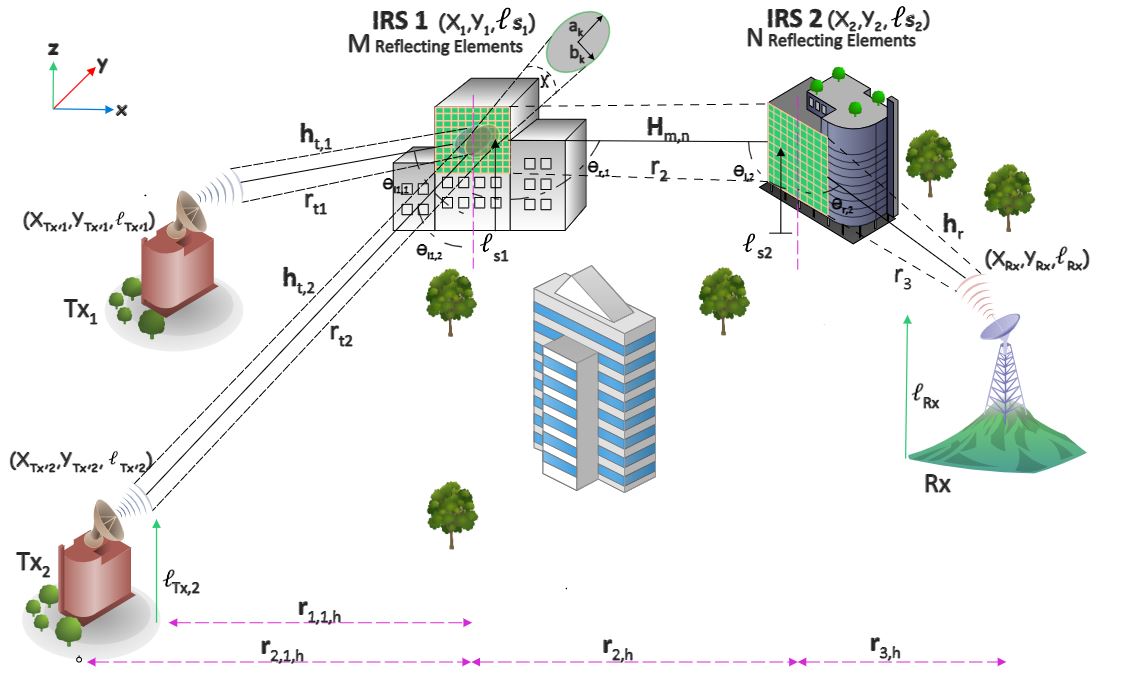}
    \caption{Cascaded IRS system model.}
    \label{fig:Cascaded IRS system model}
\end{figure*}

\section{System And Channel Model}
In our system model, we assume that we have two single-antenna users operating in an uplink multi-hop IRS communication system as shown in Fig.~\ref{fig:Cascaded IRS system model}, where the operating frequency is within the THz range. The two users are equipped with highly directional parabolic antennas and transmit the signal focused to the center of IRS$_1$. The antenna diameter for each transmitter is $D_t$ and for the receiver is $D_r$. The distances between the two users and IRS$_1$, IRS$_1$ and IRS$_2$, and between IRS$_2$ and $R_x$, are denoted as $r_{tk}$, $r_{2}$, and $r_{3}$ respectively. The horizontal distances between the transmitters and the center of IRS$_1$, the center of IRS$_1$ and the center of IRS$_2$, IRS$_2$, and the receiver $R_X$ are denoted as $r_{k,1,h}$, $r_{2,h}$, and $r_{3,h}$ respectively. The incident and reflected angles with respect to the center of the illuminated areas at IRS$_1$ and IRS$_2$ are represented as $\theta_{i1,1}, \theta_{i2,1}$, $\theta_{r,1}$, $\theta_{i,2}$, and $\theta_{r,2}$ respectively. The heights of the two transmitters, IRS$_1$, IRS$_2$, and the receiver $R_x$ are denoted as $\ell_{Tx,1}, \ell_{Tx,2}$, $\ell_{s1}$, $\ell_{s2}$, and $\ell_{Rx}$ respectively. The IRSs act as beamformers that focus the incident signal at a particular reflection direction by modifying the phases of the reflecting units (RUs). The number of RUs in IRS$_1$ is $M = M_x \times M_y$ and the number of RUs in IRS$_2$ is $N = N_x \times N_y$.\\

The transmitted signal for each user $k$, where $k \in {1,2} $, is represented by the following equation:

\begin{align} x_k = \sqrt{P_t} z_k, \end{align}

\noindent where $z_k$ represents the signal for user $k$ with unit power (i.e., $\mathbb{E}[|z_k|^2] = 1$, $\mathbb{E}[.]$ denotes the expectation), and $P_t$ represents the transmit power for each user.\\

Thus, the received signal for each user $k$ is denoted by:

\begin{align}
y_k & = \mathbf{h}_{r}^H \mathbf{\Phi}_N \mathbf{H}_{m,n}^H \mathbf{\Phi}_M \mathbf{h}^H_{t,k} x_k + n_0,\\ 
\label{eq: Received Signal at user K}
y_k & = \mathbf{h}_{r}^H \mathbf{\Phi}_N \mathbf{H}_{m,n}^H \mathbf{\Phi}_M \mathbf{h}^H_{t,k} \sqrt{P_t} z_k + n_0,\nonumber
\end{align}

\noindent where $ \mathbf{h}_{t,k}$ $\in \mathbb{C}^{1 \times M} $ is the channel between each user $k$ and IRS$_1$, $\mathbf{H}_{m,n} \in \mathbb{C}^{M \times N} $ is the channel between IRS$_1$ and IRS$_2$, $  \mathbf{h}_{r} \in \mathbb{C}^{N \times 1} $ is the channel between IRS$_2$ and the receiver, $ \large  \mathbf{\Phi}_M = \text{diag}(e^{-j\eta_1},e^{-j\eta_2},...,e^{-j\eta_M}) $ and $ \large  \mathbf{\Phi}_N = \text{diag}(e^{-j\psi_1},e^{-j\psi_2},...,e^{-j\psi_N}) $ are the phase shift reflection matrices for IRS$_1$ and IRS$_2$ respectively that satisfy the constant modulus constraint $ |\phi_m|^2 = |e^{-j\eta_m}|^2 = 1 $,  $ \forall m \in  \{1,2,...,M\}  $ and $ |\phi_n|^2 = |e^{-j\psi_n}|^2 = 1$, $\forall n \in \{1,2,...,N\}$, and $\text{diag}(.)$ denotes the diagonal matrix. Moreover, the phase shifts of the $ m^{th} $ and $ n^{th} $ reflecting elements are represented by $ \eta_m $ and $ \psi_n $, where $ \eta_m $ and $ \psi_n $  values are between $0$ and $ 2\pi $, and the noise $ n_0 \sim \mathcal{C} \mathcal{N} (0,\sigma^2) $ represents the AWGN for each user in linear scale. 

Moreover, the deterministic phase shifts corresponding to the traveled distances of the signals from each user $k$ over the first hop, and the IRS2-$R_X$ link over the third hop are represented as follows: 
\vspace{0mm}
\begin{align} \Omega_k = 2 \pi r_{t_k}/\lambda, \nonumber \end{align}  
\begin{align} \Omega_3 = 2 \pi r_{3} / \lambda . \end{align}  

\noindent where $\lambda$ is the wavelength, and it is equal to $c/f$ where c is the speed of light equal to $3 \times 10^8$  m/s and $f$ is the frequency measured in Hertz (Hz).

The transmitter and receiver channels $ \mathbf{h}_{t,k} $, and $  \mathbf{h}_{r} $ follow the Rician fading model \cite{IEEEhowto:11}, \cite{IEEEhowto:13}:

\begin{align}
\mathbf{h}_{t,k}  = \sqrt{\frac{K_1}{K_1+1}} \mathbf{\Bar h}_{t,k} + \sqrt{\frac{1}{K_1+1}} \mathbf{\tilde h}_{t,k},
\end{align}

\begin{align}
\mathbf{h}_{r} = \sqrt{\frac{K_2}{K_2+1}} \mathbf{\Bar h}_{r} + \sqrt{\frac{1}{K_2+1}} \mathbf{\tilde h}_{r},
\end{align}

\noindent where $K_1$ is the rician factor of $\mathbf{h}_{t,k}$. $\mathbf{\Bar h}_{t,k}$ $\in {C}^{1 \times M} $, and $\mathbf{\tilde h}_{t,k}$ $\in {C}^{1 \times M} $ are the LOS component, and non-LOS (NLOS) component respectively. Similarly, $K_2$ is the rician factor of $ \mathbf{h}_{r}$, $ \mathbf{\Bar h_{r}} \in {C}^{N \times 1} $ and $ \mathbf{\tilde h}_{r} \in {C}^{N \times 1} $ are the LOS component and NLOS component respectively. The channel between IRS$_1$ and IRS$_2$, $ \mathbf{H}_{m,n}  \sim \mathcal{C} \mathcal{N} (0,\mathbf{R)} $, follows the spatially correlated Rayleigh fading channel model, where $\mathbf{R}$ is the covariance matrix obtained based on the exponential spatial correlation model. It is controlled using the parameter $\rho \in [0,1] $ which represents the correlation-coefficient among the adjacent RUs, and it is expressed as: 

\begin{align}
[\mathbf{R}]_{m,n} = \rho^{|m-n|} e^{|m-n|\theta_{i,2}} 
\end{align}

\noindent where $\theta_{i,2}$ is the angle of arrival between IRS$_1$ and IRS$_2$. High values of $\rho$, result in high correlation among ${\bf{H_{mn}}}$ elements, and in cases where $\rho$ is less than $1$ (i.e. not equal to $1$), the significant correlations are between adjacent RUs only, with considerably low correlation at large distances. Further, we assume that the channels $ \mathbf{h}_{t,k}$, and $ \mathbf{h}_{r}$ are perfectly known for all the transmitters and the receiver. Even-though the channel estimation, and finding out the channel state information (CSI) \cite{IEEEhowto:14} is a challenging task for IRS-based communication networks, some studies proposed significant methods for obtaining CSIs. The authors, in \cite{IEEEhowto:15} proposed an efficient channel estimation algorithm for a double IRS-based assisted multi-user MIMO communication system to obtain the cascaded CSI. In \cite{IEEEhowto:16}, the authors conducted a comprehensive survey on channel estimation for IRS-assisted wireless communications focused on solutions that tackle practical design problems. The study in \cite{IEEEhowto:17} proposed a framework for IRS channel estimation where a small number of IRS elements are capable of processing the received signal to facilitate channel estimation. Thus, the IRS estimates the channel between itself and the BS, and between itself and the users based on the pilot signals received by the IRS semi-passive elements utilizing compressed sensing methods.






Further, the gains for the users' and receiver antennas $G_t(o)$ and $G_r(o)$ are expressed as:
\begin{equation}
    G_{\varkappa,k} (o)= 4 e_\varkappa\frac{J_1(\frac{\pi D_\varkappa \sin(o)}{\lambda})}{\sin(o)}, \varkappa\in\{t,r\}.
\end{equation} 

\noindent where $J_1(.)$ is the first-order Bessel function of the
first kind, $D_\varkappa$ is the diameter of the antenna, and $\varkappa \in \{t,r\}$ represents Tx or Rx antennas, respectively. The angle measured from the broadside of the antenna is represented as $o$ \cite{IEEEhowto:18}. 
 
Thus, the maximum gain is for $o = 0 $ and it is denoted as:

\begin{equation}
    G_{\varkappa,k}(o)= e_\varkappa (\frac{\pi D_\varkappa}{\lambda})^2, \varkappa\in\{t,r\}.
\end{equation} 
 \\
\noindent where $e_t$ and $e_r$ represents the aperture efficiencies for the $T_x$ and $R_x$ respectively. Further, the gain of every RU is expressed as \cite{IEEEhowto:18}

\begin{equation}
    G(\theta_{i1,k})=4\cos(\theta_{i1,k}), \quad 0 \leq \theta_{i1,k} \leq \pi/2,
\end{equation}

\noindent where $\theta_{i1,k}$ is the angle of incidence from user $k$ to IRS$_1$. \cite{IEEEhowto:18}.

The total losses and gains on the path between each $T_{x_k}$ and the $R_x$ are denoted by $L_{\tau,k}$. This includes the antenna gains, free space path loss (FSPL), and THz absorption loss (AS). 

\begin{align}
L_{\tau,k} =  L_{FSPL,\tau,k} ~ L_{abs,\tau,k},
\end{align}

\noindent where $L_{abs,\tau,k}$ represents the total THz absorption losses for each $T_{x_k}$. These THz losses are obtained under standard atmospheric conditions utilizing the simplified model suggested in \cite{IEEEhowto:19} , and $L_{FSPL,\tau,k}$ represents the total FSPL for each $T_{x_k}$, and it is denoted as:


\begin{align}
L_{FSPL,\tau,k} = L_{FSPL,k}~ L_{FSPL,r}.
\end{align}

\noindent $L_{FSPL,k}$ for the signal reflected from IRS$_1$ towards IRS$_2$ is represented as 

\begin{align} L_{FSPL,k} = \frac{({\frac{\lambda}{4 \pi})^2 ~ G_{t,k} G_{\theta_{i_{1,k}}} G_{\theta_{r,1}}}} {r_{t_k^2}}, \end{align} 

\noindent and $L_{FSPL,r}$ between IRS$_1$ and the receiver is expressed as 

\begin{align} L_{FSPL,r} = \frac{({\frac{\lambda}{4 \pi})^4 ~ G_{\theta_{i,2}} G_{\theta_{r,2}} G_{r}}} {r_2^2 r_3^2}. \end{align}  \\

Thus, the total FSPL for each $T_{x_k}$ is expressed as

\begin{align}\label{FSPL,total}
    L_{FSPL,\tau,k} = \bigg(\frac{\lambda}{4\pi}\bigg)^6~\frac{ G_{t,k}G(\theta_{i1,k})G(\theta_{r,1})G(\theta_{i,2}) G(\theta_{r,2}) G_{r} }{r_{tk}^2 r_{2}^2 r_{3}^2}.
\end{align}
\\










 
\section{Maximizing the Rate of a Desired User Under Interference}
\label{End-to-End Rate Analysis}
In this section, we provide the analytical derivations of the first objective, which is maximizing the rate of a desired user, while the other user is considered an interferer. Our objective is to find the optimum phases of the cascaded IRSs that maximize the received rate of the desired user. We will show that the rate maximization problem is non-convex and finding a closed-form expression of the IRS phases is mathematically intractable. Then we propose sub-optimal solution to the problem through maximizing the received power of the desired user. In addition, we propose a DDPG algorithm that maximizes rate of a the desired user.

\subsection{Rate of the Desired User Under Interference}
The analytical form of the rate of the desired user under the interference scenario can be written as 
\begin{align} \large R_k = \log_2(1+\gamma_k), \label{eq:Rate} \end{align}
\noindent where $\gamma_k$ is the signal to interference plus noise ratio (SINR) of user $k$.  The SINR, $\gamma_k$, can be written as
\begin{align}
\gamma_k= \frac{P^k_{Rx} }{\sum_{\substack{i=1 \\ i\neq k}}^K P^k_{Rx} + \sigma ^2}, \label{sinr1}
\end{align}
\noindent where $P^k_{Rx} $ is the received power of user $k$ and $\sigma^2$ is the noise variance. In the following, we provide the anlytical derivations of $P^k_{Rx}$.

 \subsubsection{User's Received Power ($P^k_{Rx}$)}
 
In our scenario, both users are transmitting at IRS$_1$, covering all IRS$_1$ elements from various angles and distances. The power reflected from the $m_{th}$ RU of IRS$_1$ can be written as in \cite{IEEEhowto:18} without including the absorption losses calculation. $L_{abs,\tau,k}$ is included in $L_{\tau,k}$ in the expression for the total received power.

\begin{equation} 
P^k_{r,m} = \bigg(\frac{\lambda}{4\pi}\bigg)^2\frac{G_{t,k}G(\theta_{i1,k})G(\theta_{r,1})}{r_{tk}^2} \times |{h}_{t,km}|^2 |\alpha_m|^2 P_t,
\end{equation}

\noindent where $\alpha_m=\alpha e^{-j\eta_m}$ designates the reflection coefficient of the $m^{th}$ RU of IRS$_1$; $G_{t,k}$ represents the $T_x$ antenna gain of user $k$; $G(\theta_{i1,k})$ and $G(\theta_{r,1})$ represent the gain of RU from the incident and reflection angles, respectively; and $r_{t,k}$ is the distance between $T_{x_k}$ and RU $m$. Similarly, the reflected power from the $n^{th}$ RU of IRS$_2$ because of being illuminated by the signal reflected by the $m^{th}$ RU of IRS$_1$ is

\begin{align}\label{Prmn}
    P^k_{r,mn} = \bigg(\frac{\lambda}{4\pi}\bigg)^4\frac{G_{t,k}G(\theta_{i1,k})G(\theta_{r,1})G(\theta_{i,2}) G(\theta_{r,2}) }{r_{tk}^2 r_{2}^2}\\ \times |{h}_{t,km}|^2 |\alpha_m|^2 |H_{mn}|^2 |\alpha_n|^2 P_t,\nonumber
\end{align}

\noindent where $\alpha_n=\alpha e^{-j\psi_n}$ designates the reflection coefficient of the $n^{th}$ RU of IRS$_2$, $H_{mn}$ represents the $(m,n)$ element of the IRS$_1$-IRS$_2$ channel matrix ${\bf{H}}$. Finally, the received captured power at the $R_x$ through channel $H_{mn}$ can be written as follows

\begin{equation} 
    P^k_{rx,mn}=\bigg(\frac{\lambda}{4\pi}\bigg)^2\frac{P^k_{r,mn}}{ r_{3}^2}G_{r} |{h}_{rn}|^2.
\end{equation}

\begin{align}
    P^k_{rx,mn} = \bigg(\frac{\lambda}{4\pi}\bigg)^6\frac{G_{t,k}G(\theta_{i1,k})G(\theta_{r,1})G(\theta_{i,2}) G(\theta_{r,2}) G_{r} }{r_{tk}^2 r_{2}^2 r_{3}^2}\\ \times |{h}_{t,km}|^2 |\alpha_m|^2  |H_{mn}|^2 |\alpha_n|^2 |{h}_{rn}|^2 P_t, \nonumber
\end{align}

and the total received power for user $k$ at the receiver ($R_x$) is expressed as \cite{IEEEhowto:18}

\begin{align}  \label{Total Received Power Initial Exp}
P^k_{Rx} = | \sqrt{L_\textrm{$\tau$,k}} \sum_{m=1}^{M}\sum_{n=1}^N |{h}_{t,km}| |\alpha_m| |H_{mn}| |\alpha_n| |h_{rn}|  \\ e^{-j\left(\varphi_{t_{k_m}} + \eta_m  + \varphi_{mn}  + \psi_n +\varphi_{r_n} + \Omega_k + \Omega_3)\right)}|^2 P_t, \nonumber
\end{align}

\noindent where $\varphi_{t_{k_m}} $ is the phase for the transmitter channel for user $k$ and $m^{th} RU$, $\varphi_{mn}$ is the phase for the $\mathbf{h}_{m,n}$ channel for $m^{th}$ and $n^{th}$ RU, $\varphi_{r_n}$ is the phase for the receiver channel for $n^{th} RU$, $|\alpha_n|$ and $|\alpha_m|$ are assumed to be equal to 1, thus equation \eqref{Total Received Power Initial Exp} can be re-written as:

\begin{align}  \label{Total Received Power}
P^k_{Rx} = | \sqrt{L_\textrm{$\tau$,k}} \sum_{m=1}^{M}\sum_{n=1}^N |{h}_{t,km}||H_{mn}| |h_{rn}|  \nonumber \\ e^{-j\left(\varphi_{t_{k_m}} + \eta_m  + \varphi_{mn}  + \psi_n +\varphi_{r_n} + \Omega_k + \Omega_3)\right)}|^2 P_t. 
\end{align}

\subsubsection{Desired User's Rate Maximization Problem}
To this end, substituting (\ref{Total Received Power}) in (\ref{sinr1}) leads to 

\begin{align}  \large \gamma_k = \frac{|\sqrt{L_\textrm{$\tau$,k}}  e^{-j\Omega_3} \mathbf{h}_{r}^H \mathbf{\Phi}_N \mathbf{H}_{m,n}^H \mathbf{\Phi}_M \mathbf{h}_{t,k}^H  e^{-j\Omega_k} |^2 P_t}{ \sum_{\substack{i=1 \\ i\neq k}}^K | \sqrt{L_\textrm{$\tau$,i}}  e^{-j\Omega_3} \mathbf{h}_{r}^H \mathbf{\Phi}_N \mathbf{H}_{m,n}^H \mathbf{\Phi}_M \mathbf{h}^H_{t,i}  e^{-j\Omega_i}|^2 P_t+ \sigma ^2}. \label{eq:Signal to Noise Ratio}
\end{align} 

Therefore to maximize the rate of the desired user under interference, we have

\begin{align} 
\max_{\mathbf{\Phi}_N,\mathbf{\Phi}_M} ~ ~ &\log_{2} \left (1 + {\gamma}_k \right), \label{eq:max_SumRate0} \\
s.t.~ &\nonumber\\
\text{C1}: &|\mathbf{\phi}_m|^2 = 1 ,   \forall m \in  \{1,2,...,M\} ,\nonumber\\
\text{C2}: &|\mathbf{\phi}_n|^2 = 1 ,   \forall n \in  \{1,2,...,N\} ,\nonumber\\
\end{align}

This optimization problem in (\ref{eq:max_SumRate0}) is NP-hard problem,  and the solution is non-trivial, because of the non-convexity due to the constant modulus constraints of IRS $1$ and IRS $2$ reflecting elements \cite{IEEEhowto:20}, and it is mathematically intractable to obtain an analytical closed form expression of the optimum phases shifts of the two IRSs.  The optimal solution will be a balance point between increasing the received SNR of the desired user while decreasing the interference from the other user which are not guaranteed to be aligned sub-objectives. Hence, we find a sub-optimal solution of (\ref{eq:max_SumRate0}) through maximizing received power of the desired user only. 

\subsubsection{Sub-optimal Solutions: Maximizing the Received Power of the Desired User}

\noindent The total received power of the desired user (e.g. user $1$) can be maximized by solving the following system of equations:

\begin{equation}\label{Solve_for_Phases}
    \eta_m+\psi_n + \varphi_{t_{1_m}}  + \varphi_{mn}  +\varphi_{r_n} + \Omega_1 + \Omega_3 = \nu, \quad \forall m,n.
\end{equation}

\noindent where $\nu$ is any constant value, which means that $P^k_{Rx}$ will be maximum when $\nu$ is constant $\quad \forall m,n$. In our case, we will choose $\nu = 0$. Thus, equation \eqref{Solve_for_Phases} describes an over-determined system of equations with M + N unknowns, and $M \times N$ equations. We illustrate below equation \eqref{Solve_for_Phases} in more detail:

\begin{align}\label{Matrix_Phases}
    \eta_1 +\psi_{M+1} & = - \varphi_{t_k,1}  - \varphi_{11}  -\varphi_{r,1} - \Omega_{k,1} + \Omega_{3,1}  \nonumber, \\
    \eta_m +\psi_{M+n} & = - \varphi_{t_k,m}  - \varphi_{mn}  -\varphi_{r,n} - \Omega_{k,m} + \Omega_{3,n}  \nonumber, \\
    \eta_M +\psi_{M+N} & = - \varphi_{t_k,M}  - \varphi_{MN}  -\varphi_{r,N} - \Omega_{k,M} + \Omega_{3,N} , \nonumber \\
\end{align}


$$ {M \times N} 
\underbrace{ \begin{bmatrix}
      1 & 0 & \dots & 0 & 1 & 0 & \dots & 0 \\
      0 & 1 & \ddots & \vdots & 0 & 1 & \ddots & \vdots \\
      \vdots & \ddots &\ddots & \vdots & \vdots & \ddots &\ddots & \vdots \\
      0 & \dots & 0 & 1 &  0 & \dots & 0 & 1 \\
\end{bmatrix}}_{M + N} 
\times
\underbrace{  \begin{bmatrix}
    \eta_1 \\
    \eta_2 \\
    \vdots \\
    \eta_m \\
    \vdots \\
    \eta_{M-1} \\
    \eta_{M} \\
    \psi_{M+1} \\
    \vdots \\
    \psi_{M+n} \\
    \vdots \\
    \psi_{M+N-1} \\
    \psi_{M+N} \\
  \end{bmatrix}}_{(M + N) \times 1} 
  = \underbrace{
\begin{bmatrix}
    - \varphi_{t_k,1}  - \varphi_{11}  -\varphi_{r,1} - \Omega_{k,1} + \Omega_{3,1}   \\
    \vdots      \\  
    - \varphi_{t_k,M}  - \varphi_{M1}  -\varphi_{r,1} - \Omega_{k,M} + \Omega_{3,1}   \\ 
    \vdots   \\ 
    - \varphi_{t_k,1}  - \varphi_{12}  -\varphi_{r,2} - \Omega_{k,1} + \Omega_{3,2}   \\
    \vdots\\
    - \varphi_{t_k,M}  - \varphi_{M2}  -\varphi_{r,2} - \Omega_{k,M} + \Omega_{3,1}   \\ 
    \vdots \\
    - \varphi_{t_k,m}  - \varphi_{mn}  -\varphi_{r,n} - \Omega_{k,m} + \Omega_{3,n}   \\  
    \vdots      \\  
    - \varphi_{t_k,1}  - \varphi_{1N}  -\varphi_{r,N} - \Omega_{k,1} + \Omega_{3,N}   \\
    \vdots      \\ 
    - \varphi_{t_k,M}  - \varphi_{MN}  -\varphi_{r,N} - \Omega_{k,M} + \Omega_{3,N}    \\  
\end{bmatrix}}_{(M\times N) \times 1}  
$$
\begin{equation} 
\end{equation} 
 

This is almost always inconsistent and has no solution. However, we solve this problem using sub-optimal mathematical techniques such as pseudo inverse and block solution to obtain the unknowns $\eta_m$ and $\psi_n$, calculate $ \large  \mathbf{\Phi}_M$, $ \large  \mathbf{\Phi}_N $, and the received power for user $1$, and then we compare the results generated from these techniques to those obtained from deep reinforcement learning solution.

\paragraph{Pseudo-Inverse Solution}

The pseudo-inverse solution for the over-determined system is expressed as follows: 
 
\begin{align} \mathbf{A} \mathbf{\Theta} = \mathbf{C},\end{align} 

\noindent where $\mathbf{\Theta}$ with dimensions $(M+N)\times 1$ is the matrix that represents IRS$_1$, and IRS$_2$ phase shifts from $\eta_1$ to $\psi_{M+N}$, $\mathbf{A}$ with dimensions ${(M \times N)\times (M+N)}$ represents the binary matrix, and $\mathbf{C}$ with dimensions ${(M \times N)\times 1}$ represents the matrix containing constant values such as the phase shifts of the transmitter channel$~\textbf{h}_{t,k}^{(t)}$, the phase shifts of the channel between IRS$_1$ and IRS$_2$ ${\bf{H_{mn}}}$, and the phase shifts of the receiver channel $~\textbf{h}_{r}^{(t)}$.


\begin{align} \mathbf{A^+}  = (\mathbf{A^T} \mathbf{A})^{-1}\mathbf{A^T}, \end{align}

\noindent where $\mathbf{A^+}$  \text{is pseudo-inverse of a matrix} $\mathbf{A}$. Thus, the pseudo-inverse solution $\mathbf{\Theta}$ is expressed as

\begin{align}\mathbf{\Theta} = \mathbf{A^+}  \mathbf{C}. \end{align}

\paragraph{Block Solution} Moreover, based on the spatially correlated channel assumption, a low-complexity solution built on the exponential correlation model can be developed and is controlled using the correlation-coefficient $\rho$ between the adjacent RUs. High $\rho$ values lead to high correlation across ${\bf{H_{mn}}}$ elements. This results in two-dimensional blocks of high correlations across the diagonal elements of the channel matrix. Building on this, the SINR can be maximized by selecting one element from each block and replicating its phase response to all other elements in the same block. This solution is denoted as a block solution.

In the case where the channel correlation is very high, the channel ${\bf{H_{mn}}}$ has a block structure, where the elements in the channel matrix form groups, each having the same phase with no correlation between the contiguous blocks. Thus, the number of phases that the IRSs need to compensate for is minimized to the number of blocks in the channel matrix ${\bf{H_{mn}}}$. Thus, the over-determined system becomes solvable if the number of blocks in the channel matrix $N_{\textrm{blk}} \leq M + N $, where $N_{\textrm{blk}}$ denotes the number of blocks in the channel.

Therefore, the total received signal power when using the block solution can be rewritten as:

\begin{align}  \label{Total Received Power Block Solution}
P^k_{Rx} = | \sqrt{L_\textrm{$\tau$,k}} \sum_{v=1}^V\sum_{w=1}^W |{h}_{t,kv}||H_{vw}| |h_{rw}|  \\ e^{-j\left(\varphi_{t_{k_v}} + \eta_v  + \varphi_{vw}  + \psi_w +\varphi_{r_w} + \Omega_k + \Omega_3)\right)}|^2 P_t,\nonumber
\end{align}

\noindent where $V = \frac{M}{N_{\textrm{blk}}}$, and $W = \frac{N}{N_{\textrm{blk}}}$ represent the number of blocks, and v, and w represent the index of the blocks in the correlated channel.

\begin{algorithm}[h]
\caption{Block Solution-based Framework}
    \label{Alg:Block}
    \small{
\begin{algorithmic}[1]
\State \textbf{Input:} $h_{t}$, $M$, $H_{mn}$, $N$, $h_{t}$, $N_{\textrm{blk}}$
\State \textbf{Output:} $\eta_n$, $\psi_m$, $P_{Rx_k}$

\If{$ M + N \leq M \times N$}
    \State Divide IRS$_1$ and IRS$_2$ elements into $N_{\textrm{blk}}$ blocks.
    \If{$ M + N \geq N_{\textrm{blk}}$}
       \State calculate the total received power using \eqref{Total Received Power Block Solution}.
    \Else
       \If {$ M + N < N_{\textrm{blk}}$}
          \State Solve \eqref{Solve_for_Phases} using the pseudo-inverse solution, and calculate the total received power using \eqref{Total Received Power}.
       \EndIf  \textbf{endif}\\
    \EndIf \textbf{endif} \\
\EndIf  \textbf{endif}
\end{algorithmic}}
\end{algorithm}

\begin{prop} 
\label{proposition I}
The total received signal power for each $T_{x_k}$ at the $R_x$ in eq. \eqref{Total Received Power} can be expressed as:
\end{prop}

\begin{align} P^k_{Rx} = |\sqrt{L_\textrm{$\tau$,k}}  e^{-j\Omega_3}\mathbf{h}_{r}^H \mathbf{\Phi}_N \mathbf{H}_{mn}^H \mathbf{\Phi}_M \mathbf{h}_{t,k}^H  e^{-j\Omega_k} |^2  P_t.\end{align} 

The proof is in the Appendix \ref{Appendix I}.

\section{Maximizing the Sum Rate for Both Users}

In this section, we provide the mathematical derivations of the optimization problem of our second objective, which is maximizing the sum rate of both users. In particular, in our second objective, we find the optimum phase shifts of IRS 1 and IRS 2 elements that maximize the combined sum rate for all users at the receiver. Moreover, we derive a a loose upper bound that is used to benchmark the results of our proposed solutions. The sum rate can be written as

\begin{align}   
\large
R_{sum} =\sum_{k=1}^{K} \log_{2} \left (1 + \large \gamma_k \right).\label{eq:SumRate}
\end{align}

Accordingly, the formulated problem at IRS$_1$ and IRS$_2$ is to find out the phase shift matrices $\mathbf{\Phi}_N$, and $\mathbf{\Phi}_M$ that maximizes $ R_{sum} $, and it is expressed as:

\begin{align} 
\max_{\mathbf{\Phi}_N,\mathbf{\Phi}_M}\sum_{k=1}^{K} & \log_{2} \left (1 + {\gamma}_k \right), \label{eq:max_SumRate} \\
s.t.~ &\nonumber\\
\text{C1}: &|\mathbf{\phi}_m|^2 = 1 ,   \forall m \in  \{1,2,...,M\} ,\nonumber\\
\text{C2}: &|\mathbf{\phi}_n|^2 = 1 ,   \forall n \in  \{1,2,...,N\} ,\nonumber\\
\end{align}

Similar to the optimization problem in (\ref{eq:max_SumRate0}) this optimization problem is NP-hard problem,  and the solution is non-trivial, because of the non-convexity due to the constant modulus constraints of IRS $1$ and IRS $2$ reflecting elements \cite{IEEEhowto:20}, and it is nearly not possible to obtain an analytical solution by mathematical methods for multi-hop IRS optimization. In order to solve it, we leverage DRL technique, specifically DDPG, instead of solving the challenging problem mathematically. The details of the proposed DDPG algorithm are given in section \ref{Proposed DDPG for Cascaded IRS Phase Control}.


\subsection{Upper bound on Performance}
\label{Upper bound on Performance}

By assuming the case of null interference in (\ref{eq:Signal to Noise Ratio}), a loose upperbound \ref{eq:SumRate-Upperbound} on the sum rates can be determined as follows:

The SINR on user $k$ becomes:
\begin{align}  \large \gamma^U_k = \frac{\Bigg|\sqrt{L_\textrm{$\tau$,k}}  |\mathbf{h}_{r}^H| |\mathbf{H}_{m,n}^H| |\mathbf{h}_{t,k}^H| \Bigg|^2 P_t}{\sigma ^2}. \label{eq:SNR-Upperbound}
\end{align}  





Hence, the upper bound on sum rate can be written as: 

\begin{align}   
\large
R^U_{sum} =\sum_{k=1}^{K} \log_{2} \left (1 + \large \gamma^U_k \right).\label{eq:SumRate-Upperbound}
\end{align}


\section{Proposed DDPG for Cascaded IRS Phase Control}
\label{Proposed DDPG for Cascaded IRS Phase Control}

\begin{figure*} [t]
   \centering
    \includegraphics[width=0.9\linewidth]{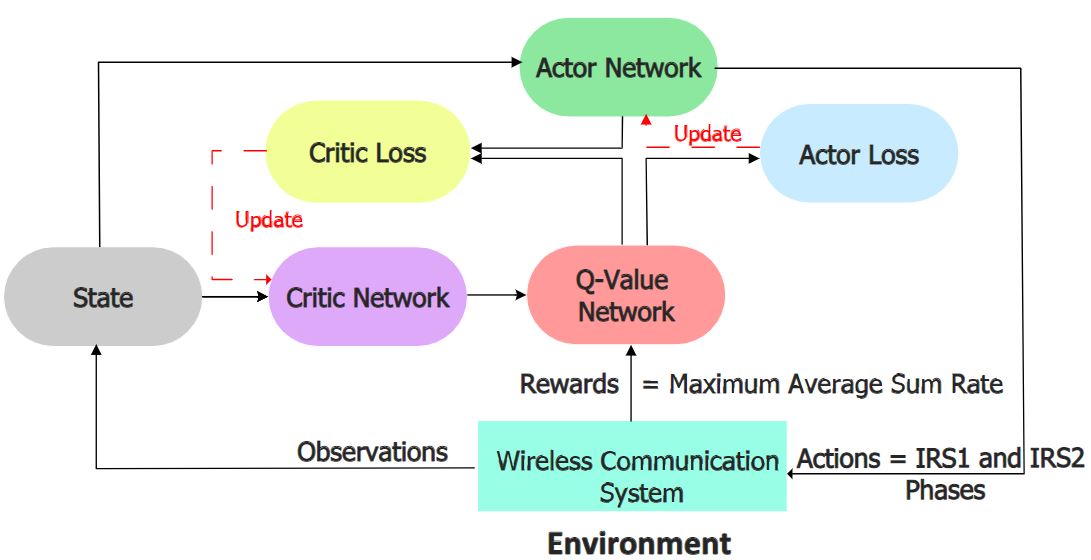} 
     \caption {DDPG model}  
     \label{fig:DDPG model}
\end{figure*}
As stated earlier, the optimization problems in (\ref{eq:max_SumRate0}) and (\ref{eq:max_SumRate}) are non-convex and finding the optimum phases that maximize the rate through exhaustive search is computationally infeasible. Hence, we design DDPG solutions to find the optimum phases of the cascaded IRS.

In this section, we introduce the scheme of the proposed DDPG algorithm in (see Fig.~\ref{fig:DDPG model}) to solve the optimization problem in \eqref{eq:max_SumRate0} and \eqref{eq:max_SumRate} for the cascaded IRS system. Deep Q-Networks are not suitable since they deal with discrete time spaces only. Moreover, the convergence of the policy gradient (PG) algorithm is not sufficient in the context of wireless communication. DDPG is a model-free reinforcement learning that merges the advantages of both the Q-networks and the PG scheme and overcomes the disadvantages of both algorithms. It utilizes both the continuous state and action spaces\cite{IEEEhowto:21}. Thus, the optimization problem in \eqref{eq:max_SumRate0} and \eqref{eq:max_SumRate} can be solved by learning the policy. 

DDPG scheme consists of several key components, which include the agents, state $\textbf{s}^{(T)}$, the action $\textbf{a}^{(T)}$, the reward  $\textbf{r}^{(T)}$, the policy function $\mu$, and the Q-value function $Q(\textbf{s},\textbf{a}|\theta^{Q})$. The agents operating in our system are the IRS$_1$ and IRS$_2$. The states $\textbf{s}^{(T)}$ in our system are the received SINR for $T_{x_1}$ at the $R_x$, the received SINR for $T_{x_2}$ at the $R_x$, and the sum rate for users at time step ${(T-1)}$.

The actions $\textbf{a}^{(T)}$ are the phases of IRS$_1$ and IRS$_2$, and the reward $\textbf{r}^{(T)}$ is the received power for user $1$ for our first objective, and the users' sum rate for second objective. Our goal is to optimize the average rewards as this involves instant and future rewards. The DDPG scheme is composed of four neural networks, the critic network, the actor network, the target actor, and the target critic networks to ensure stability.

\subsection{DDPG System Mapping}

The initial step in solving the optimization problem of the system model is to map it into the fundamental elements of the DDPG algorithm, specifically, the state space, the action space, and the reward function. Below we discuss the details of this mapping in addition to the general behavior of the DDPG algorithm.

\subsubsection{State space}

The state space of the DDPG agent at timestep ${(T)}$ is represented as follows:

For the first objective:
 
\begin{align}
    \textbf{s}^{(T)}=[\large \gamma_1^{(T-1)},\large \gamma_2^{(T-1)},P_{Rx_k}^{(T-1)}],
\end{align}

For the second objective:

\begin{align}
    \textbf{s}^{(T)}=[\large \gamma_1^{(T-1)},\large \gamma_2^{(T-1)},R_{sum}^{(T-1)}],
\end{align}


\noindent where $\large \gamma_1^{(T-1)}$, $\large \gamma_2^{(T-1)}$, $P_{Rx_k}^{(T-1)}$, and $R_{sum}^{(T-1)}$ represent the received SINR for $T_{x_1}$ at the $R_x$, the received SINR for $T_{x_2}$ at the $R_x$, the received power for $T_{x_1}$ at the $R_x$, and the sum rate for users at time step ${(T-1)}$ respectively.

\subsubsection{Action Space}

The actions are the IRS$_1$ and IRS$_2$ phase shift values. The output is an array that defines the phase of each element in the IRS. Thus, the action space is defined by the following policy function:

\begin{align}
    \textbf{a}^{(T)}=\mu(\textbf{s}^{(T)}|\theta^\mu)+\textbf{n}{(T)}
\end{align}
\noindent where $\mu$ is defined as the policy function and $\theta^\mu$ represents parameters,  weights of neural network, and $\textbf{n}(T)$ is the Ornstein Uhlenbeck (OU) process-based action noise~\cite{IEEEhowto:22}.

Because action space is continuous, the exploration of the action space is handled with the noise that is generated by the OU process. The OU process samples the noise from a correlated normal distribution.

\subsubsection{Reward function}

The reward function for the first objective is defined based on the maximum received power for the desired user:

\begin{align} 
\label{eqreward}
    r^{(T)} = P_{Rx_k}^{(T)}
\end{align}

\noindent where $P_{Rx_k}^{(T)}$ is the maximum received power for user $1$.

For the second objective, it is defined as the maximum sum rate for users :

\begin{align} 
\label{eqreward}
    r^{(T)} = R_{sum}^{(T)}
\end{align}

\noindent where $R_{sum}^{(T)}$ is the actual sum-rate for the users.

\subsubsection{DDPG Algorithm Framework}

The objective of the DDPG algorithm is to train the agents IRS$_1$ and IRS$_2$ to take actions that maximize the long-term average reward (i.e. user $1$' received power and user's sum rate) throughout the changes of the environments. The agents IRS$_1$ and IRS$_2$ adjust the randomized policy and the phases shift matrices in a way that deals with the random environmental statistical behavior to maintain a long-term average reward, rather than an instantaneous response to the channel random changes. 

For each iteration, the agents IRS$_1$ and IRS$_2$ observe the state which includes the received SINR of the previous state for $T_{x1}$, $\large \gamma_1^{(T-1)}$, at the $R_x$, the received SINR of the previous state for $T_{x2}$, $\large \gamma_2^{(T-1)}$, at the $R_x$, and the reward of the previous state. After that, it calculates the action $\large \mathbf{\Phi}_M$ and $\large \mathbf{\Phi}_N$ that maximizes the long-term reward. This is accomplished by the actor network, whereas the critic network accepts the state and the action as inputs and outputs the expected reward (i.e. user $1$' received power and user's sum rate). After the reward calculation, a new state is observed, and the agents IRS$_1$ and IRS$_2$ will modify the phases accordingly till the system learns how to reach the optimal reward. To increase stability, the target actor and critic networks are updated periodically based on the newest actor and critic parameter values. 

As revealed in algorithm~\ref{Alg:DDPG}, in step~1 we initialize the replay buffer $D$ of the agent with capacity $C$. Then we initialize the weights of the actor network and critic network in step~2. In step~3, the actor target network and the critic target network are initialized. The phase shifts for all elements of IRS$_1$ and IRS$_2$ are chosen randomly from 0 to $2\pi$ at the start of each episode. The following steps from step~4 to step~13 are repeated for each timestep (i.e.iteration). For each timestep, we obtain the following channels $~\textbf{h}_{t,k}^{(T)},~\textbf{h}_{m,n}^{(T)},~\textbf{h}_{r}^{(T)}$, $\large \gamma_1^{(T-1)}$, $\large \gamma_2^{(T-1)}$, and $R_{sum}^{(T-1)}$, observe the state $\textbf{s}^{(T)}$ for the agents IRS$_1$ and IRS$_2$, and the actor network will determine an action $\textbf{a}^{(T)}$ (i.e. Phase shift matrices)  with exploration noise (OU) in step~5. The action $\textbf{a}^{(T)}$ is reformed into a phase shift matrices for IRS$_1$ and IRS$_2$ $\large  \mathbf{\Phi}_M = \text{diag}(e^{-j\eta_1},e^{-j\eta_2},...,e^{-j\eta_M}) $ and $ \large  \mathbf{\Phi}_N = \text{diag}(e^{-j\psi_1},e^{-j\psi_2},...,e^{-j\psi_N})$. After the agents determine and execute the action, a reward $\textbf{r}^{(T)}$ is calculated, and a new state $\textbf{s}^{(T+1)}$ is observed. The state $\textbf{s}^{(T)}$, action $\textbf{a}^{(T)}$, reward $\textbf{r}^{(T)}$, and the new state $\textbf{s}^{(T+1)}$ are stored as one transition in the replay memory $D$ in steps~6 and 7. Then, the critic network samples a random mini-batch of transitions from the main memory in step~8 to calculate the target Q-value $\Tilde{Q}(\textbf{s}^{(i)},\textbf{a}^{(i)}|\theta^{Q'})$ in step~9 using Bellman equation. In step~10, the weights of the target actor network are updated by minimizing the loss using the obtained target Q-value. In step~11, the weights of the target critic network are updated according to the sampled policy gradient. Finally, the target actor and critic networks are updated using the soft updates $\tau$ to increase the learning stability as shown in step~12. 

\subsubsection{Neural Network Architecture}

The architecture of the DDPG scheme is composed of $4$ neural networks which include the actor and critic networks as well as the target actor and target critic networks which are used to improve the stability of the learning process. The target networks are used in the Q-target formula to estimate the value of the future states that are used to train the current networks.


\begin{algorithm}[h]
\caption{DDPG-based Framework}
    \label{Alg:DDPG}
    \small{
\begin{algorithmic}[1]
\State \textbf{Initialization:} Set $T=0$ and initialize reply buffer of DDPG agent $\mathcal{D}$ with capacity M.
\State Randomly initializes the weights of actor networks $\theta^{\mu}$ and critic networks $\theta^Q$.
\State Initialize target networks: $\theta^{\mu'}\leftarrow\theta^{\mu}$ and $\theta^{Q'}\leftarrow\theta^Q$. 
\For{$T=1$ to $\infty$}
\State Observe state $\textbf{s}^{(T)}$ and select an action with exploration OU noise $\textbf{a}^{(T)}=\mu(\textbf{s}^{(T)}|\theta^\mu)+\textbf{n}_T$
\State Execute action $\textbf{a}^{(T)}$ at IRS$_1$ and IRS$_2$.  
\State Receive the immediate reward $r^{(T)}$, and observe next state $s^{(T+1)}$, store transition $(\textbf{s}^{(T)},\textbf{a}^{(T)},r^{(T)},\textbf{s}^{(T+1)})$ in $D$.
\State Randomly sample mini-batch transitions from $\mathcal{D}$:

$B\leftarrow\{(\textbf{s}^{(i)},\textbf{a}^{(i)},r^{(i)},\textbf{s}^{(i+1)})\} \in \mathcal{D}$ .
\State Compute the targets:

$\Tilde{Q}(\textbf{s}^{(i)},\textbf{a}^{(i)}|\theta^{Q'})=r^{(i)}+\Gamma Q(\textbf{s}^{(i+1)},\mu(\textbf{s}^{(i)}|\theta^{\mu'})|\theta^{Q'})$
\State Update the $\theta^Q$ in the critic network by minimizing the loss:

$L = \frac{1}{|B|}\sum_{i=1}^{|B|}\big(\Tilde{Q}(\textbf{s}^{(i)},\textbf{a}^{(i)}|\theta^{Q'})-Q(\textbf{s}^{(i)},\textbf{a}^{(i)}|\theta^Q)\big)^2$
\State Update the $\theta^\mu$ in actor network according to the sampled policy gradient:

$\nabla_{\theta\mu}\boldsymbol{J}\approx \frac{1}{|B|}\sum_{i=1}^{|B|}\nabla_a{Q}(\textbf{s}^{(i)},\textbf{a}^{(i)}|\theta^{Q})\nabla_{\theta\mu}\mu(\textbf{s}^{(i)}|\theta^\mu)$

\State Update the target networks:

$\theta^{Q'}\leftarrow\tau\theta^Q+(1-\tau)\theta^{Q'}$

$\theta^{\mu'}\leftarrow\tau\theta^\mu+(1-\tau)\theta^{\mu'}$
    \EndFor
    \State \textbf{end for}
\end{algorithmic}}
\end{algorithm}

\subsection{Complexity Analysis}

To reveal the complexity of the DRL algorithm, we demonstrate a quantitative analysis of the proposed DRL-based algorithm $\mathcal{C_{DRL}}$ versus the complexity of pseudo-inverse, block solution, and exhaustive search methods. The DRL algorithm is a neural network-based algorithm and its architecture possesses the multi-layer perceptron (MLP) structure, which is a fully connected class of feed-forward artificial neural network (ANN). Thus, the complexity of the forward pass in MLP is a vector or matrix multiplication. For our system, the evaluation of the DRL complexity depends mainly on the calculations throughout the exploitation phase. Hence, we are interested in the complexity of the trained network (steady-state) which relies heavily on the actor network (i.e. forward network architecture).

Deep neural networks are composed of multiple layers, an input layer, an output layer, and hidden layers. We assume that $S$ is the number of states, which is the size of the actor network's input, $Ui$ is the number of neurons in each layer's input, $n$ is the number of hidden layers, $Uj$ is the number of neurons in each layer's output, $A$ is the number of actions, which is the size of the actor network's output. Thus, the complexity of the input layer is $O$ ($S$ $\times$ $Ui$), the complexity of the hidden layers is $O$  ($n$ $\times$ $Ui$ $\times$ $Uj$), and the complexity of the output layer is $O$  ($Uj$ $\times$ $A$). Hence, the overall complexity of the DRL algorithm is \begin{align} \mathcal{C_{DRL}} = S \times Ui + n \times Ui \times Uj + Uj \times A. \end{align} 

Further, the DRL algorithm will always select the action $A$ that yields the highest reward, and performs a linear search on the output. Therefore, the overall computational complexity of  one forward pass in the neural network is expressed as \cite{IEEEhowto:23}: 

\begin{align} \mathcal{C_{DRL}} = S \times Ui + n \times Ui \times Uj + Uj \times A + A. \end{align} 


On the other hand, the pseudo-inverse solution of a matrix $\mathcal{A}$ with size $M N \times (M + N) $ can be calculated from its singular value decomposition (SVD) as $O( (M N)^2 \times (M + N) )$, where $M N > (M + N) $, and $M$ and $N$ are the number of elements in IRS$_1$ and IRS$_2$ respectively \cite{IEEEhowto:24}, \cite{IEEEhowto:25}. However, for the block solution, the complexity of a matrix $\mathcal{A}$ with size $\frac{M N}{N^2_{\textrm{blk}}} \times \frac{M + N}{N_{\textrm{blk}}} $ is reduced to $O( (\frac{M N}{N^2_{\textrm{blk}}})^2 \times (\frac{M + N}{N_{\textrm{blk}}}))$, where $N_{\textrm{blk}} < (M + N)$.

Moreover, the complexity of the exhaustive scheme $ \mathcal{C_{ES}}$ assuming $K$ users, $M$ elements for IRS$_1$, and $N$ elements for IRS$_2$, and phase search steps $\Xi= \lfloor\frac{2\pi}{\Delta \Phi}\rfloor$, can be expressed as 
\begin{align}
\mathcal{C_{ES}} =O(K\times (\Xi+1)^{(M+N)}).
\end{align} 

Thus, the complexity of the DRL algorithm is much lower than that of the pseudo-inverse, block solution, and exhaustive search methods as the number of IRS elements increases as shown in the following figures Fig.~\ref{fig:Complexity}, and Fig.~\ref{fig:Complexity2}. 

\begin{figure}
    \centering
    \includegraphics[width=0.6\linewidth]{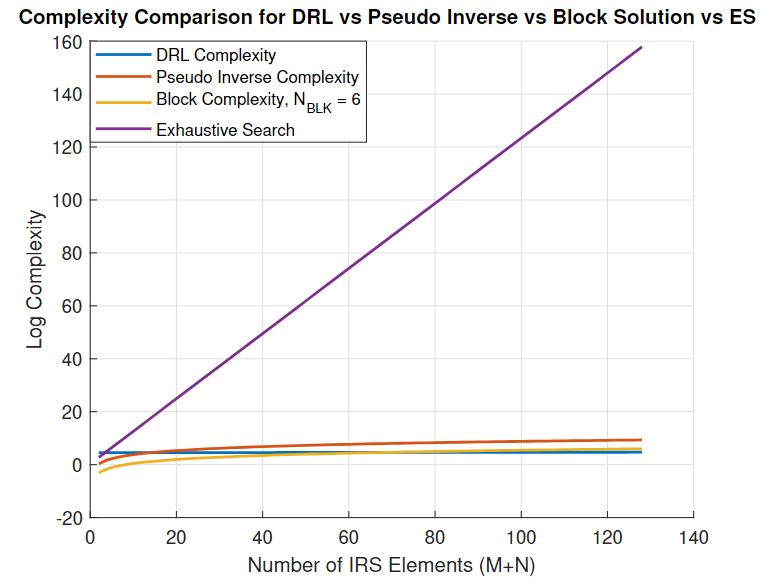}
    \caption{Complexity of DRL vs. Pseudo Inverse vs. Block Solution. M=N=64, NBLK = 6.}
    \label{fig:Complexity}
\end{figure}
\vspace{0mm}
\begin{figure}
    \centering
    \includegraphics[width=0.6\linewidth]{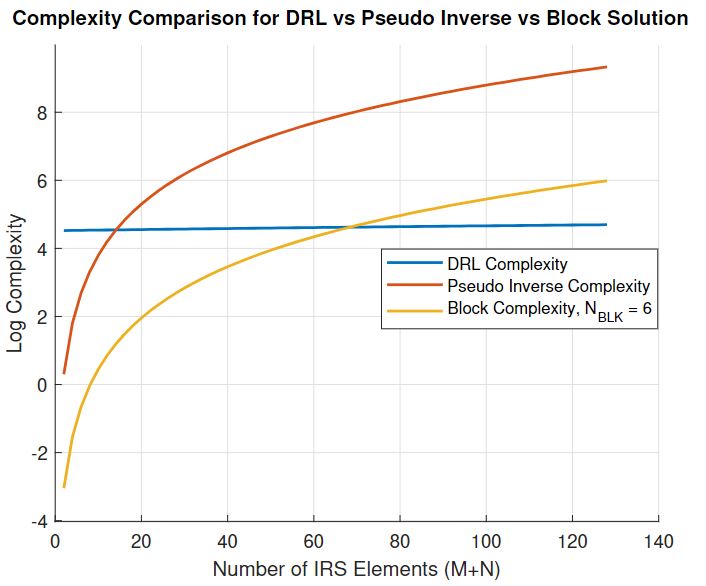}
    \caption{Complexity of DRL vs. Pseudo Inverse vs. Block Solution vs. Exhaustive Search. M=N=64, NBLK = 6.}
    \label{fig:Complexity2}
\end{figure}
\vspace{0mm}



\section{Simulation Results}

        
        
 

In this section, we evaluate the performance of the proposed DDPG-based cascaded  IRS-assisted wireless THz communications. To reveal the performance of our DDGP system, we need to compare it with a benchmark scheme considering the scenario of maximizing the rate for the desired user, and the scenario of maximizing the sum rate for both users. Therefore, when maximizing the rate for the desired user (i.e. user $1$), we provide two benchmarking schemes as reference models for our system with reflecting elements $M=18$ for IRS$_1$, and $N=18$ for IRS$_2$; the first scheme is based on pseudo-inverse, and the second is based on block solution. Further, when maximizing the sum rate for both users, we compare the sum rates obtained from using the DDPG algorithm to a discretized exhaustive search approximation to prove that our DDPG scheme performance is close to the exhaustive search. We employ this procedure to find out the optimum phase shift matrices that result in an approximation to the maximum sum rate. To avoid the high complexity of the exhaustive search algorithm, we consider a limited number of reflecting elements $M=4$ for IRS$_1$, and $N=4$ for IRS$_2$. For each IRS$_1$ and IRS$_2$ element, we consider the phase shifts between 0 and $2\pi$ with a search step of $2\pi/72$. This will give us $(72+1)^{M+N}$ combinations of phase shift reflection matrices. After obtaining the optimum phase shift matrices, we calculate the sum rates accordingly for users. Moreover, we compare the DDPG sum rates to those calculated based on random phase generation (i.e. without optimization) as another way to benchmark our system.

 
The default parameters used in the simulation for the DDPG-based cascaded IRS algorithm are shown in Table \ref{table:2}. The number of reflecting elements used for IRS$_1$ and IRS$_2$ is $18$. The number of users is $K = 2$, the number of antennas per user is $ N_t = 1$ the number of antennas at the receiver is $ N_r =1 $, and the wavelength is $\lambda = 10^{-3}$. The channel between user $1$ and IRS$_1$, user $2$ and IRS$_1$, IRS$_2$, and the receiver follows the rician fading model with rician factor $K1$ = $K2$ = $10$. The path loss exponent for the channel between the transmitters and IRS$_1$ is $2$, and the path loss exponent for the channel between IRS$_2$ and the receiver is $2$. The carrier frequency is $f = 300 \times 10^9$, the bandwidth is $BW = 2 \times 10^9$, the noise spectral density $N_{PSD} = -174$ db/Hz, and the noise figure at the receiver $F_{dB} = 10$ db. Moreover, the coordinates for IRS$_1$ is ($x_{r1} = 5$,$y_{r1} = 10$,$h_{r1} = 12$), the coordinates for IRS$_2$ is ($x_{r2} = 10$,$y_{r2} = 10$,$h_{r2} = 12$), the distance between user $1$ and IRS$_1$ is variable with range between $r_{t1} = 3$ m and $15$ m, the distance between user $2$ and IRS$_1$ is $15$ m, and the reflection coefficients of IRS$_1$ and  IRS$_2$ is $\alpha = 1$. The coordinates of the receiver $R_x$ is ($x_{rx} = 20$,$y_{rx} = 0$,$h_{r} = 5$). The antenna diameter is $D_t = 0.12$ m. The height of user $1$ and user $2$ is $h_t=5$. We define the distance ratio as the distance $r_{t1}$ between user $1$ and IRS$_1$, divided by the distance $r_{t2}$ between user $2$ and IRS$_1$. Numerical results for the sum rates are calculated using $10^3$ Monte-Carlo simulations. 


\renewcommand\thetable{2}
\begin{table} [t]
\caption{Parameters Used in Simulation} 
\centering 
\begin{tabular}{| c | c |} 
\hline 
\textbf{Simulation Parameters} & \textbf{Values}\\ 
\hline
Number of Users ($K$) & 2 \\ 
\hline
Number of antennas per user $ N_t $ & 1 \\
\hline
Number of antennas at the receiver $ N_r $ & 1 \\
\hline
Speed of the light $c$ & $3 \times 10 ^8$ \\
\hline
Carrier Frequency $ f $ & $300 \times 10 ^9$ \\
\hline
Wavelength $ \lambda $ & $1 \times 10 ^{-3}$ \\
\hline 
Number of IRS$_1$ Reflecting Elements (M) & $18$ \\ 
\hline
Number of IRS$_2$ Reflecting Elements (N) & $18$ \\ 
\hline
X-axis of IRS$_1$ $x_{r1}$ & $5$ \\
\hline
Y-axis of IRS$_1$ $y_{r1}$ & $10$ \\ 
\hline
X-axis of IRS$_2$ $x_{r2}$ & $10$ \\
\hline
Y-axis of IRS$_2$ $y_{r2}$ & $10$ \\ 
\hline
height of IRS$_1$ $h_{r1}$ & $12$ \\ 
\hline
height of IRS$_2$ $h_{r2}$ & $12$ \\ 
\hline
Distance between User $1$ and IRS$_1$ & $3$  to $15$  \\ 
\hline
Distance between User $2$ and IRS$_1$ & $15$  \\ 
\hline
Heights of User $1$ and User $2$ $h_t$ & $5$ \\ 
\hline
IRS$_1$ and IRS$_2$ Reflection Coefficients $\alpha$ & $1$ \\ 
\hline
IRS$_1$ and IRS$_2$ half-power Spacing $d_x$ & $\lambda/2$ \\ 
\hline
IRS$_1$ and IRS$_2$ Element Spacing $d_y$ & $\lambda/2$ \\ 
\hline
Antenna diameter in meters $D_t$ & $0.12$ \\ 
\hline
X-axis of Rx $x_{rx}$ & $20$ \\
\hline
Y-axis of Rx $y_{rx}$ & $0$ \\ 
\hline
height of Rx $h_r$ & $5$ \\ 
\hline
Bandwidth & $2 \times 10 ^9$ MHz \\
\hline
Noise power spectral density $N_{PSD}$ & $-174$ dB/Hz\\  
\hline
Noise figure at the receiver $F_{dB}$ & $10$ \\  
\hline
Average Noise power in dB $N0$ & $-174$ dB/Hz\\  
\hline
Noise power in linear scale $ no $ & $7.9621 \times 10 ^{-11}$ \\
\hline
Transmitters to IRS$_1$ Path loss exponent & $2$ \\ 
\hline
IRS$_2$ to receiver $R_x$ Path loss exponent & $2$ \\
\hline
Rician Factor & $10$ \\  
\hline
Critic Network learning rate & $3 \times 10 ^{-4}$ \\
\hline
Actor Network learning rate & $1 \times 10 ^{-4}$ \\
\hline
Target Critic Network learning rate & $3 \times 10 ^{-4}$ \\
\hline
Target Actor Network  learning rate & $1 \times 10 ^{-4}$ \\
\hline
Discount factor of the future reward $\Gamma$ & $0.99$ \\
\hline
Coefficient of Soft Updates $\tau$ & $1 \times 10 ^{-3}$ \\
\hline
Batch size & $128$ \\  
\hline
Replay Buffer Capacity $ \mathcal{C} $ \ & $10^5$\\ 
\hline
Number of episodes & $10000$ \\  
\hline
\end{tabular}
\label{table:2} 
\end{table} 

The proposed DDPG scheme is composed of actor and critic networks. Both networks are dense neural networks (DNN) with 4 layers. Each layer is nn.Linear and accepts two parameters, the first is the input size and the second is the output size. For the actor network, the states are the input with a size equal to 3 neurons, and the output is the action with size of 36 neurons. In addition to the input and output layers, there are two hidden layers each that accepts 128 neurons as input and outputs 128 neurons, followed by the ReLU activation function. To provide enough gradient, the output layer of the actor network uses the tanh(·). For the critic network, the input consists of the number of states and the number of actions, both are concatenated to represent the input of the critic network. This is followed by two hidden layers between the input layer and the output layer, each that accepts 128 neurons as input, and outputs 128 neurons, followed by the ReLU activation function. The output layer of the critic network represents the Q-value with 36 neurons. To update the parameters, both actor and critic networks use Adam optimizer. Furthermore, the results are generated by considering the average rate for over 1000 iterations. The learning rate of the actor network is set to $3 \times 10^{-4}$, and the learning rate of the critic network is set to $1 \times 10^{-4}$. The discount factor of the future reward $\Gamma$, the batch size is equal to $128$, the replay buffer $\mathcal{C}$ is equal to $10^5$, and the number of episodes is 10000.

\begin{figure} 
   \centering
    \includegraphics[width=0.6\linewidth]{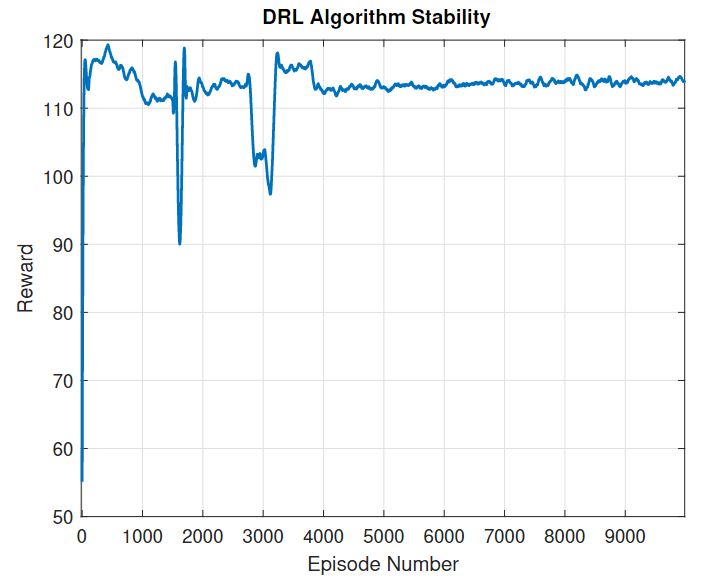} 
     \caption {DDPG algorithm convergence. Rewards vs Episodes.}  
     \label{fig:DRL Convergence}
\end{figure}
\vspace{0mm}

The generated results shown in Fig.~\ref{fig:DRL Convergence} reveal the convergence of the DDPG algorithm. The figure shows the rewards versus the episodes, and that the rewards are increasing with time. This reveals that the learning process is conducted successfully.

\subsection{Simulation Results for Maximizing The Received Power At The Receiver $R_x$}

\begin{figure}
     \centering
     \begin{subfigure}[b]{0.5\textwidth}
         \centering
         \includegraphics[width=1\textwidth]{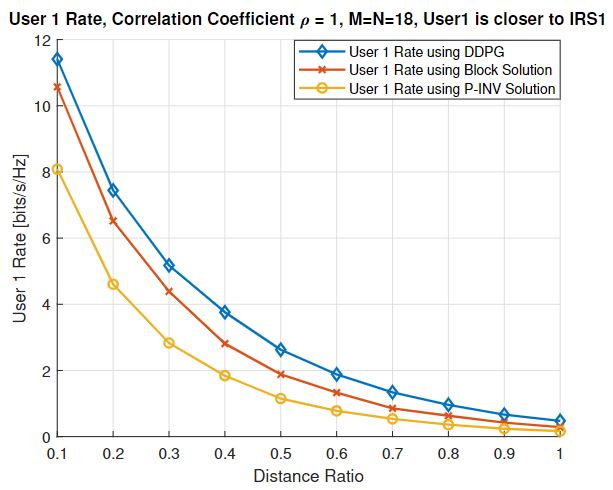}
         \caption{Correlation-Factor $\rho$ equal to 1.0.}
         \label{fig:Rate1_vs_DR_CF_1.0}
     \end{subfigure}
     \hfill
     \begin{subfigure}[b]{0.5\textwidth}
         \centering
         \includegraphics[width=1\textwidth]{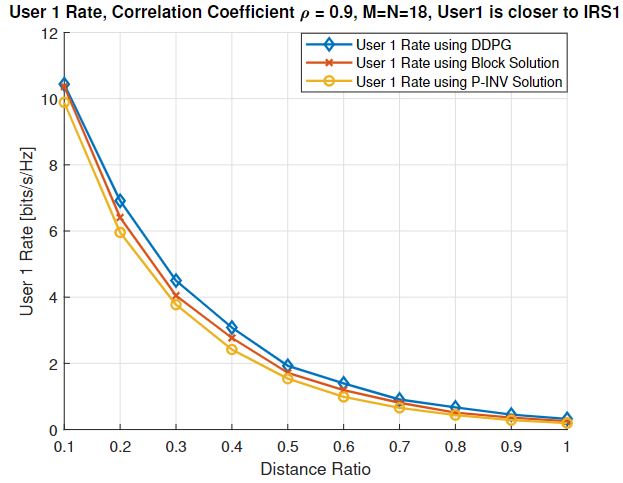}
         \caption{Correlation-Factor  $\rho$ equal to 0.9.}
         \label{fig:Rate1_vs_DR_CF_0.9}
     \end{subfigure}
     \hfill
     \begin{subfigure}[b]{0.5\textwidth}
         \centering
         \includegraphics[width=1\textwidth]{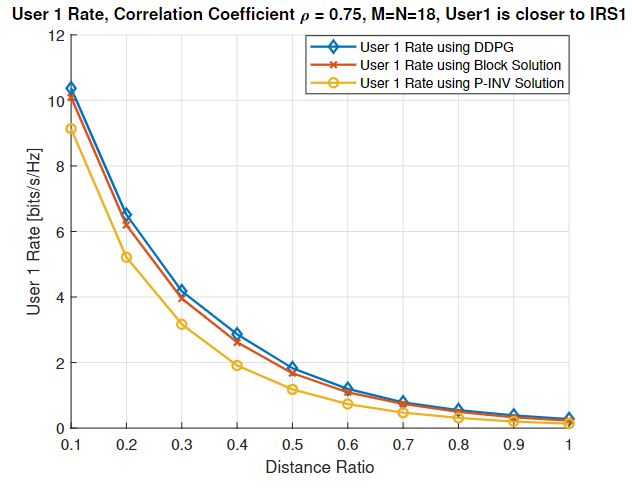}
         \caption{Correlation-Factor  $\rho$ equal to 0.75.}
         \label{fig:Rate1_vs_DR_CF_0.75}
     \end{subfigure}
       \caption{User 1's Rate  vs Distance ratio}
       \label{fig:User 1's Rate vs Distance ratio}  
\end{figure}
\vspace{0mm}
\begin{figure}
     \centering
     \begin{subfigure}[b]{0.5\textwidth}
         \centering
         \includegraphics[width=1\textwidth]{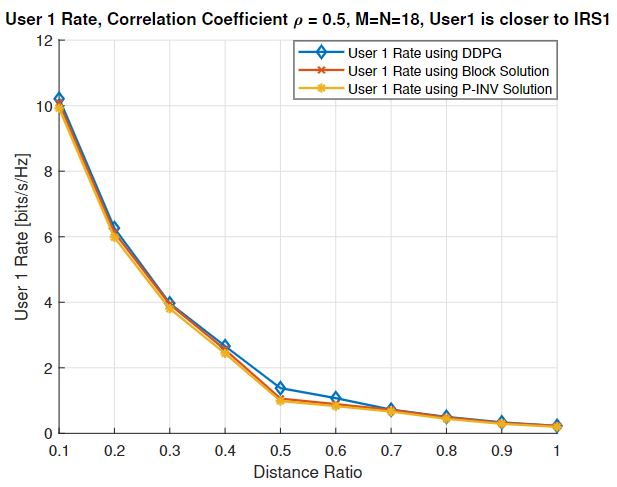}
         \caption{Correlation-Factor  $\rho$ equal to 0.5.}
         \label{fig:Rate1_vs_DR_CF_0.5}
     \end{subfigure}
     \hfill
     \begin{subfigure}[b]{0.5\textwidth}
         \centering
         \includegraphics[width=1\textwidth]{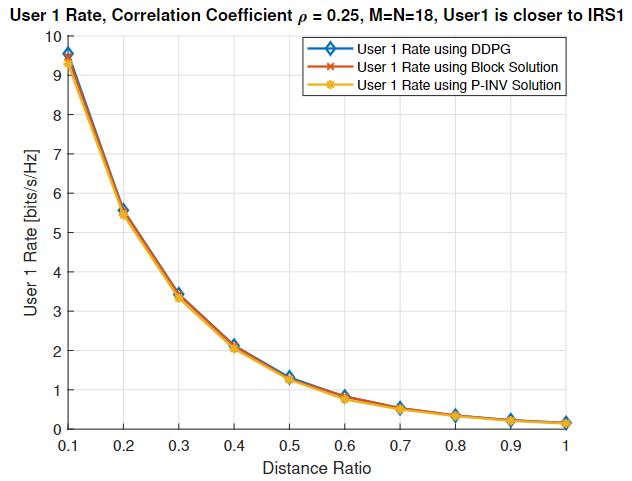}
         \caption{Correlation-Factor  $\rho$ equal to 0.25.}
         \label{fig:Rate1_vs_DR_CF_0.25}
     \end{subfigure}
       \caption{User 1's Rate  vs Distance ratio}
       \label{fig:User 1's Rate vs Distance ratio P2}  
\end{figure}
\vspace{0mm}
In the following simulations, we demonstrate the results for maximizing the rate for the desired user (i.e. user $1$) at the receiver by plotting the rate of user $1$ versus the distance ratio between user $1$ and user $2$ utilizing various schemes such as DDPG, block solution, and pseudo-inverse solution methods.



Fig.~\ref{fig:User 1's Rate  vs Distance ratio} and Fig.~\ref{fig:User 1's Rate vs Distance ratio P2} shows user $1$'s rate for the DDPG scheme versus the distance ratio between $0.2$ and $1$. As the distance ratio increases, the data rate for user $1$ decreases, because user $2$ will be closer to user $1$, and thus the interference from user $2$ to user $1$ increases. Moreover, Fig.~\ref{fig:Rate1_vs_DR_CF_1.0} shows that user $1$'s rate for the DDPG scheme exceeds that of pseudo-inverse and block-solution methods with a correlation-coefficient $\rho$ equal to 1.0. This reveals the effectiveness of the DDPG algorithm and that it is superior to pseudo-inverse and block-solution. 





Furthermore, the following figures Fig.~\ref{fig:Rate1_vs_DR_CF_0.9} Fig.~\ref{fig:Rate1_vs_DR_CF_0.75}, Fig.~\ref{fig:Rate1_vs_DR_CF_0.5}, and Fig.~\ref{fig:Rate1_vs_DR_CF_0.25} demonstrate that when the correlation-coefficient $\rho$ decreases, user's $1$ rate decreases for DDPG algorithm, block-solution, and pseudo-inverse solution. This shows the importance of correlation channels in our scenario and their benefit in increasing the data rate. On top of that, the DDPG scheme still achieves higher rates than block-solution and the pseudo-inverse solution even when the correlation-coefficient $\rho$ decreases. On the other hand, it is important to note that, for low values of correlation-coefficient $\rho$ the difference between the DDPG's data rates and other schemes retracts.

\begin{figure} 
   \centering
    \includegraphics[width=0.6\linewidth]{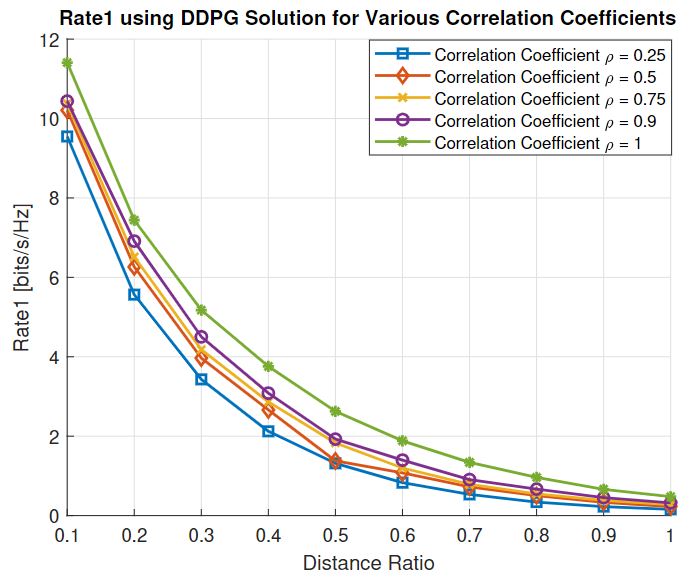} 
     \caption {User 1's Rate for DDPG  vs Distance ratio for Various correlation-coefficients $\rho$. M=N=18.}  
     \label{fig:Rate1_DDPG_Various_CorrFactors}
\end{figure}
\vspace{0mm}


Fig.~\ref{fig:Rate1_DDPG_Various_CorrFactors} demonstrate the rates for the DDPG scheme versus the distance-ratio for various correlation-coefficients. It is clear that when the correlation-coefficient $\rho$ increases the data rates for the DDPG scheme increase, since increasing the value of correlation, will increase the learning efficiency of the DDPG algorithm. Thus, the DDPG scheme achieves higher data rates than other methods especially when the correlation-coefficient $\rho$ is high. 
 
\subsection{Simulation Results for Maximizing the Sum Rate for Both Users}

\begin{figure} 
   \centering
    \includegraphics[width=0.6\linewidth]{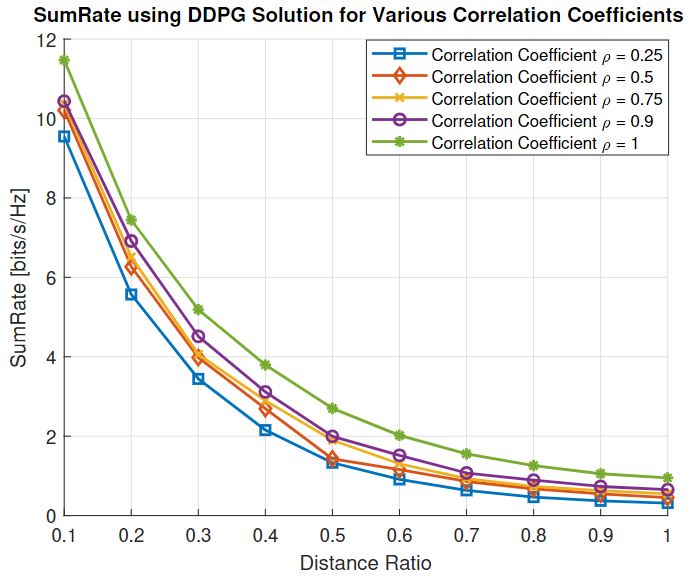} 
     \caption {DDPG Sum Rate vs Distance ratio for Various correlation-coefficients $\rho$. M=N=18.}  
     \label{fig:SumRate_DDPG_Various_CorrFactors}
\end{figure}

In the following simulations, we show the results for maximizing the sum rate for both users at the receiver. We plot the sum rate versus the distance ratio between user $1$ and user $2$ utilizing the DDPG method for various correlation coefficients as shown in Fig.~\ref{fig:SumRate_DDPG_Various_CorrFactors}. It is obvious from the results that the sum rates obtained from the DDPG solution increase with the increase of the correlation-coefficient $\rho$. 

\begin{figure} 
   \centering
    \includegraphics[width=0.6\linewidth]{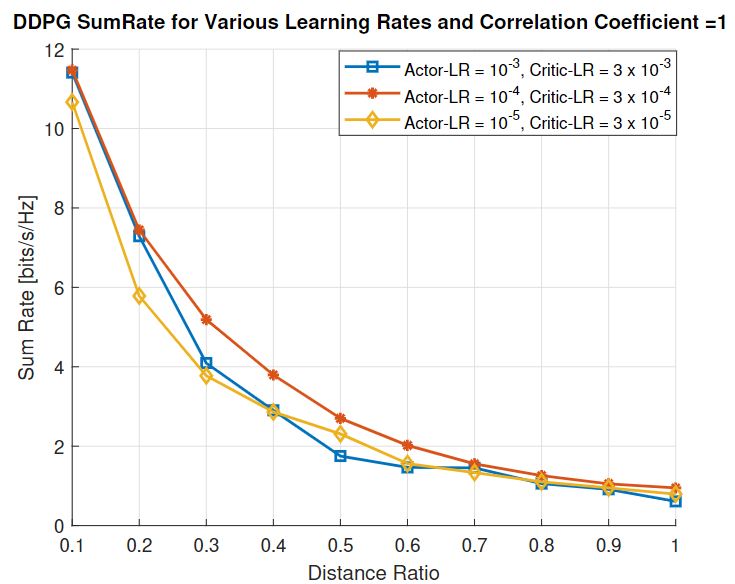} 
     \caption {DDPG Sum Rate vs Distance ratio for Various learning rates and correlation-coefficient $\rho$ = 1.0. M=N=18.}  
     \label{fig:Sum-rate_DDPG_Various_Learning_Rates}
\end{figure}

Further, in all our simulations, constant learning rates were used for our proposed DDPG scheme, which is $10^{-4}$ for actor network, and $3 \times 10^{-4}$ for the critic network. The influence of the learning rate on the DDPG data rates is shown in Fig.~\ref{fig:Sum-rate_DDPG_Various_Learning_Rates}, which reveals a comparison between different learning rates, i.e., {$10^{-3}$, $10^{-4}$, $10^{-5}$} for actor network, and {$3 \times 10^{-3}$, $3 \times 10^{-4}$, $3 \times 10^{-5}$} for critic network. Thus, the highest DDPG rate is achieved when the actor networks' learning rate equals $10^{-4}$ and the critic networks' learning rate equals $3 \times 10^{-4}$. Therefore, the average rewards are determined by the learning rates (e.g. $10^{-4}$). Too small learning rates $10^{-5}$or too large learning rates $10^{-3}$ produce lower average rewards, where the learning rate $10^{-4}$ achieves better rewards.

\vspace{0mm}
\vspace{0mm}

\begin{figure} 
   \centering
    \includegraphics[width=0.6\linewidth]{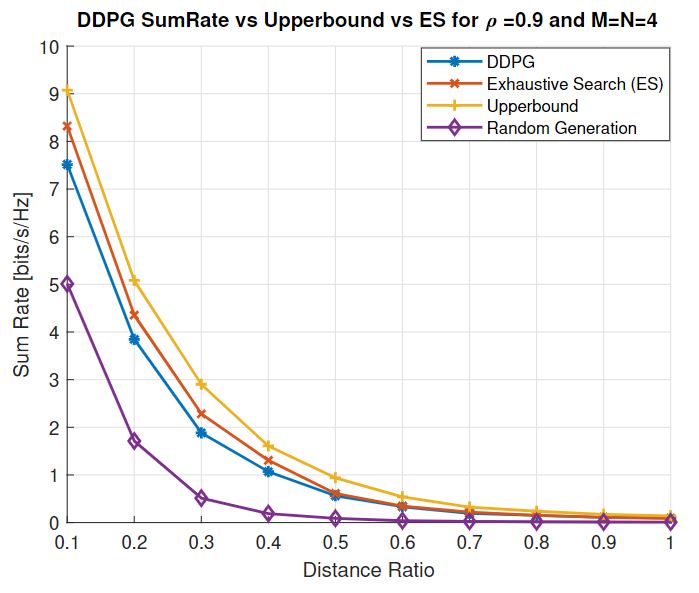} 
     \caption {Comparison between DDPG Sum Rate, exhaustive search, and upper bound vs Distance Ratio for Various correlation-coefficients $\rho$ = 0.9. M=N=4.} 
     \label{fig:DDPG_Vs_Upper_Bound_M_N_4}
\end{figure}

To verify the performance of our DDPG scheme, we compare the sum rates produced by the DDPG algorithm to the discretized exhaustive search algorithm that is used to calculate the maximum sum rate by obtaining the optimum phase shift matrix. The complexity of the exhaustive search is too much high so the number of reflecting elements used for IRS$_1$ is $M$ = 4, and for IRS$_2$ is $N$ = 4 rather than $M$ = $N$ = 18. For each IRS element, we take into consideration the phases between 0 and $2\pi$ with a search step size equal to $\frac{2\pi}{72}$. Thus the number of combinations for the phase shift matrices is equal to $(72+1)^4$. The sum rates are calculated for two users and $100$ Monte-Carlo simulations. The results are revealed in Fig.~\ref{fig:DDPG_Vs_Upper_Bound_M_N_4} where the DDPG algorithm sum rates are close to the exhaustive search with the specified granularity.

\vspace{0mm}

\section{Conclusion} 

In this paper, we considered the uplink multiple access scenario of the cascaded IRS system to combat the short-range communications in THz networks, intending to achieve two objectives. The first objective is to maximize the rate of the desired user. We showed that the problem is non-convex and finding a closed form expression is mathematically intractable. Therefore, we proposed two sub-optimal solution for maximizing the received power of the desired user. The second objective is to maximize the sum rate for both users which is also non-convex problem and more complicated. We employed the DDPG algorithms which are capable of coping with non-convex optimization problems to solve the maximization problem for the cascaded IRS system. DDPG algorithm obtains the optimum IRS phases that maximize the received rate for the desired user, and the sum rate for both users. Simulation results for the first objective reveal that DDPG can achieve higher data rates than sub-optimal methods, pseudo-inverse and block-solution. For the second objective, it is clear that the DDPG sum rates are close to the discretized exhaustive search with a search step equal to $2*pi/72$. Further, DDPG reveals the significance of the correlation in the channels to enhance the learning process and achieve higher data rates.

\onecolumn
\appendix
\label{Appendix I}
\subsection{Proof of Proposition ~\ref{proposition I} (See page~\pageref{proposition I})} 
 
The total received signal power for each $T_{x_k}$ at the $R_x$ in eq. \eqref{Total Received Power} can be expressed as: 

\begin{align}  P^k_{Rx} = |\sqrt{L_\textrm{$\tau$,k}}  e^{-j\Omega_3}\mathbf{h}_{r}^H \mathbf{\Phi}_N \mathbf{H}_{m,n}^H \mathbf{\Phi}_M \mathbf{h}_{t,k}^H  e^{-j\Omega_k} |^2  P_t. \nonumber 
\end{align}  
\begin{proof} 

\begin{itemize}

\item Multiply the receiver channel $\mathbf{h}_{r}^H $ by IRS$_2$ phase shift reflection matrix $\mathbf{\Phi}_N$:
 
$$\begin{bmatrix} \mathbf{h}_{r}^H \mathbf{\Phi}_N \end{bmatrix} =  \underbrace{\begin{bmatrix} {h}_{r1}^*,\dots,{h}_{rn}^* ,\dots,{h}_{rN}^* \end{bmatrix}}_{1 \times N} \times 
\underbrace{
\begin{bmatrix}
      e^{-j\psi_1} & 0 & \dots & \dots & \dots & 0 \\
      0 & e^{-j\psi_2} & 0 & \ddots & \ddots& 0 \\
      \vdots & 0 & \ddots & \ddots & \ddots & \vdots\\
      \vdots& \ddots & \ddots & e^{-j\psi_n} & \ddots & \vdots \\
      \vdots & \ddots& \ddots & \ddots & \ddots & \vdots \\
      0 & \dots & \dots & \dots & 0 & e^{-j\psi_N} \\
\end{bmatrix}}_{N \times N}$$ 
$$\begin{bmatrix} \mathbf{h}_{r}^H \mathbf{\Phi}_N \end{bmatrix}  = \underbrace{
\begin{bmatrix}
 {h}_{r1}^* e^{-j\psi_1} ,\dots,{h}_{rn}^* e^{-j\psi_n} ,\dots,{h}_{rN}^* e^{-j\psi_N}
\end{bmatrix}}_{1 \times N}$$
\item  Multiply the result of the previous operation by $\mathbf{h}_{m,n}^H$:
 
$$\begin{bmatrix} \mathbf{h}_{r}^H \mathbf{\Phi}_N \mathbf{H}_{mn}^H \end{bmatrix} = \underbrace{ \begin{bmatrix} {h}_{r1}^* e^{-j\psi_1} ,\dots,{h}_{rn}^* e^{-j\psi_n} ,\dots,{h}_{rN}^* e^{-j\psi_N}  \end{bmatrix}}_{1 \times N}  \times 
\underbrace{
\begin{bmatrix}
      h_{11}^* & h_{21}^* & \dots & \dots & \dots & h_{M1}^*\\
      h_{12}^* & h_{22}^* & \ddots & \ddots & \ddots& h_{M2}^* \\
      \vdots & \ddots & \ddots & \ddots & \ddots & \vdots\\
      \vdots& \ddots & \ddots & H_{mn}^* & \ddots & \vdots \\
      \vdots & \ddots& \ddots & \ddots & \ddots & \vdots \\
       h_{1N}^* & \dots & \dots & \dots &\dots & h_{MN}^* \\
\end{bmatrix}}_{N \times M} 
$$
$$\begin{bmatrix} \mathbf{h}_{r}^H \mathbf{\Phi}_N \mathbf{H}_{mn}^H \end{bmatrix} = 
\underbrace{
\begin{bmatrix}
 \sum_{n=1}^{N} {h}_{rn}^* e^{-j\psi_n} {h}_{1n}^* ,\sum_{n=1}^{N} {h}_{rn}^* e^{-j\psi_n} {h}_{2n}^* ,\dots,\sum_{n=1}^{N} {h}_{rn}^* e^{-j\psi_n} {h}_{MN}^*
\end{bmatrix}}_{1 \times M}
$$\\
\item Multiply the result of the previous operation by IRS$_1$ phase shift reflection matrix $\mathbf{\Phi}_M$:

$$ \begin{bmatrix} \mathbf{h}_{r}^H \mathbf{\Phi}_N \mathbf{H}_{mn}^H \mathbf{\Phi}_M\end{bmatrix} = $$
$$\underbrace{
\begin{bmatrix}
 \sum_{n=1}^{N} {h}_{rn}^* e^{-j\psi_n} {h}_{1n}^* ,\sum_{n=1}^{N} {h}_{rn}^* e^{-j\psi_n} {h}_{2n}^* ,\dots,\sum_{n=1}^{N} {h}_{rn}^* e^{-j\psi_n} {H}_{MN}^*
\end{bmatrix}}_{1 \times M} \times \underbrace{
\begin{bmatrix}
      e^{-j\eta_1} & 0 & \dots & \dots & \dots & 0 \\
      0 & e^{-j\eta_2} & 0 & \ddots & \ddots& 0 \\
      \vdots & 0 & \ddots & \ddots & \ddots & \vdots\\
      \vdots& \ddots & \ddots & e^{-j\eta_m} & \ddots & \vdots \\
      \vdots & \ddots& \ddots & \ddots & \ddots & \vdots \\
      0 & \dots & \dots & \dots & 0 & e^{-j\eta_M} \\
\end{bmatrix}}_{M \times M}    
$$
$$ \begin{bmatrix} \mathbf{h}_{r}^H \mathbf{\Phi}_N \mathbf{H}_{mn}^H \mathbf{\Phi}_M\end{bmatrix} = 
\underbrace{
\begin{bmatrix}
 \sum_{m=1}^{M} \sum_{n=1}^{N} {h}_{rn}^* e^{-j\psi_n} {H}_{mn}^* e^{-j\eta_m}
\end{bmatrix}}_{1 \times M}
$$
\item  Multiply the result of the previous operation by the transmitter channel of user $K$ $\mathbf{h}_{t,k}^H$:
 
$$ \begin{bmatrix}  \mathbf{h}_{r}^H \mathbf{\Phi}_N \mathbf{H}_{mn}^H \mathbf{\Phi}_M \mathbf{h}_{t,k}^H \end{bmatrix} = \underbrace{
\begin{bmatrix}
 \sum_{m=1}^{M} \sum_{n=1}^{N} {h}_{rn}^* e^{-j\psi_n} {H}_{mn}^* e^{-j\eta_m}
\end{bmatrix}}_{1 \times M}  \times  \underbrace{
\begin{bmatrix}
  {h}_{t,k,1}^* \\  {h}_{t,k,2}^* \\ \vdots \\ {h}_{t,k,M}^*    
\end{bmatrix}}_{M \times 1}
$$

\begin{align} \mathbf{h}_{r}^H \mathbf{\Phi}_N \mathbf{H}_{mn}^H \mathbf{\Phi}_M \mathbf{h}_{t,k}^H  \nonumber =  \sum_{m=1}^{M} \sum_{n=1}^{N} {h}_{rn}^* e^{-j\psi_n} {H}_{mn}^* e^{-j\eta_m} {h}_{t,km}^*, \end{align}

\begin{align}\mathbf{h}_{r}^H \mathbf{\Phi}_N \mathbf{H}_{m,n}^H \mathbf{\Phi}_M \mathbf{h}_{t,k}^H \nonumber =  \sum_{m=1}^{M} \sum_{n=1}^{N} |{h}_{rn}| e^{-j\phi_{rn}} e^{-j\psi_n} |{H}_{mn}| e^{-j\phi_{mn}} e^{-j\eta_m} |{h}_{t,km}| e^{-j\phi_{t,km}},  \end{align}
\begin{align}  \mathbf{h}_{r}^H \mathbf{\Phi}_N \mathbf{H}_{mn}^H \mathbf{\Phi}_M \mathbf{h}_{t,k}^H  \nonumber =  \sum_{m=1}^{M} \sum_{n=1}^{N} |{h}_{t,km}| |{H}_{mn}| |{h}_{rn}| e^{-j(\phi_{t,km} +e^{-j\eta_m} + e^{-j\phi_{mn}} + e^{-j\psi_n} + \phi_{rn})}, \end{align}
\begin{align}   \mathbf{h}_{r}^H \mathbf{\Phi}_N \mathbf{H}_{mn}^H \mathbf{\Phi}_M \mathbf{h}_{t,k}^H  \nonumber =  \sum_{m=1}^{M} \sum_{n=1}^{N} |{h}_{t,km}||\alpha_m| |{H}_{mn}| |\alpha_n| |{h}_{rn}| e^{-j(\phi_{t,km} +\eta_m + \phi_{mn} + \psi_n + \phi_{rn})},  \end{align}

\item From the obtained result we can deduce that the total received signal power for each $T_{x_k}$ at the $R_x$ can be written as follows:
\vspace{0mm}
\begin{align}  
P^k_{Rx} & \nonumber = |\sqrt{L_\textrm{$\tau$,k}} e^{-j\Omega_3}\mathbf{h}_{r}^H \mathbf{\Phi}_N \mathbf{H}_{mn}^H \mathbf{\Phi}_M \mathbf{h}_{t,k}^H  e^{-j\Omega_k} |^2  P_t,  \\ & \nonumber = | \sqrt{L_\textrm{$\tau$,k}} \sum_{m=1}^{M}\sum_{n=1}^N |{h}_{t,km}| |\alpha_m| |H_{mn}| |\alpha_n| |h_{rn}|  e^{-j\left(\varphi_{t_{k_m}} + \eta_m  + \varphi_{mn}  + \psi_n +\varphi_{r_n} + \Omega_k + \Omega_3)\right)}|^2 P_t, 
\end{align}

\end {itemize}

\end{proof}
\twocolumn


\begin{thebibliography} {1} 

\bibitem{IEEEhowto:1} C. Chaccour, M. N. Soorki, W. Saad, M. Bennis, P. Popovski and M. Debbah, "Seven Defining Features of Terahertz (THz) Wireless Systems: A Fellowship of Communication and Sensing," in IEEE Communications Surveys and Tutorials, DOI: 10.1109/COMST.2022.3143454. 

\bibitem{IEEEhowto:2} R. Imran, M. Odeh, N. Zorba and C. Verikoukis, "Quality of Experience for Spatial Cognitive Systems within Multiple Antenna Scenarios," in IEEE Transactions on Wireless Communications, vol. 12, no. 8, pp. 4153-4161, August 2013, doi: 10.1109/TWC.2013.071113.122037.

\bibitem{IEEEhowto:3} I. Yildirim, A. Uyrus and E. Basar, "Modeling and Analysis of Reconfigurable Intelligent Surfaces for Indoor and Outdoor Applications in Future Wireless Networks," in IEEE Transactions on Communications, vol. 69, no. 2, pp. 1290-1301, Feb. 2021, doi: 10.1109/TCOMM.2020.3035391.

\bibitem{IEEEhowto:4} Z. Chen, X. Ma, C. Han and Q. Wen, "Towards intelligent reflecting surface empowered 6G terahertz communications: A survey," in China Communications, vol. 18, no. 5, pp. 93-119, May 2021, DOI: 10.23919/JCC.2021.05.007.

\bibitem{IEEEhowto:5} T. V. Nguyen, T. P. Truong, T. M. T. Nguyen, W. Noh, and S. Cho, "Achievable Rate Analysis of Two-Hop Interference Channel with Coordinated IRS Relay," in IEEE Transactions on Wireless Communications, DOI: 10.1109/TWC.2022.3154372.

\bibitem{IEEEhowto:6} W. Mei and R. Zhang, "Multi-Beam Multi-Hop Routing for Intelligent Reflecting Surfaces Aided Massive MIMO," in IEEE Transactions on Wireless Communications, vol. 21, no. 3, pp. 1897-1912, March 2022, DOI: 10.1109/TWC.2021.3108020.

\bibitem{IEEEhowto:7} Q. Sun, P. Qian, W. Duan, J. Zhang, J. Wang and K. -K. Wong, "Ergodic Rate Analysis and IRS Configuration for Multi-IRS Dual-Hop DF Relaying Systems," in IEEE Communications Letters, vol. 25, no. 10, pp. 3224-3228, Oct. 2021, DOI: 10.1109/LCOMM.2021.3100347.

\bibitem{IEEEhowto:8} Z. Zhang and Z. Zhao, "Weighted Sum-Rate Maximization for Multi-Hop RIS-Aided Multi-User Communications: A Minorization-Maximization Approach," 2021 IEEE 22nd International Workshop on Signal Processing Advances in Wireless Communications (SPAWC), 2021, pp. 106-110, DOI: 10.1109/SPAWC51858.2021.9593114.

\bibitem{IEEEhowto:9} A. Almohamad, M. Hasna, N. Zorba, and T. Khattab, "Performance of THz Communications Over Cascaded RISs: A Practical Solution to the Over-Determined Formulation," in IEEE Communications Letters, vol. 26, no. 2, pp. 291-295, Feb. 2022, DOI: 10.1109/LCOMM.2021.3132655.

\bibitem{IEEEhowto:10} C. Huang et al., "Hybrid Beamforming for RIS-Empowered Multi-hop Terahertz Communications: A DRL-based Method," 2020 IEEE Globecom Workshops (GC Wkshps, Taipei, Taiwan, 2020, pp. 1-6, doi: 10.1109/GCWkshps50303.2020.9367503.

\bibitem{IEEEhowto:11} C. Huang et al., "Multi-Hop RIS-Empowered Terahertz Communications: A DRL-Based Hybrid Beamforming Design," in IEEE Journal on Selected Areas in Communications, vol. 39, no. 6, pp. 1663-1677, June 2021, DOI: 10.1109/JSAC.2021.3071836.

\bibitem{IEEEhowto:12} C. Soni and N. Gupta, "Channel Estimation of Spatial Correlated Channel in Massive MIMO," 2021 8th International Conference on Computing for Sustainable Global Development (INDIACom), New Delhi, India, 2021, pp. 836-841.

\bibitem{IEEEhowto:13} K. Feng, Q. Wang, X. Li and C. -K. Wen, "Deep Reinforcement Learning Based Intelligent Reflecting Surface Optimization for MISO Communication Systems," in IEEE Wireless Communications Letters, vol. 9, no. 5, pp. 745-749, May 2020, doi: 10.1109/LWC.2020.2969167.

\bibitem{IEEEhowto:14} N. Zorba and A. I. Perez-Neira, "Opportunistic Grassmannian Beamforming for Multiuser and Multiantenna Downlink Communications," in IEEE Transactions on Wireless Communications, vol. 7, no. 4, pp. 1174-1178, April 2008, doi: 10.1109/TWC.2008.060972.

\bibitem{IEEEhowto:15} B. Zheng, C. You and R. Zhang, "Efficient Channel Estimation for Double-IRS Aided Multi-User MIMO System," in IEEE Transactions on Communications, vol. 69, no. 6, pp. 3818-3832, June 2021, doi: 10.1109/TCOMM.2021.3064947.

\bibitem{IEEEhowto:16} B. Zheng, C. You, W. Mei, and R. Zhang, "A Survey on Channel Estimation and Practical Passive Beamforming Design for Intelligent Reflecting Surface Aided Wireless Communications," in IEEE Communications Surveys and Tutorials, vol. 24, no. 2, pp. 1035-1071, Second quarter 2022, doi: 10.1109/COMST.2022.3155305. 

\bibitem{IEEEhowto:17} X. Hu, R. Zhang and C. Zhong, "Semi-Passive Elements Assisted Channel Estimation for Intelligent Reflecting Surface-Aided Communications," in IEEE Transactions on Wireless Communications, vol. 21, no. 2, pp. 1132-1142, Feb. 2022, doi: 10.1109/TWC.2021.3102446.

\bibitem{IEEEhowto:18} K. Ntontin et al., "Reconfigurable Intelligent Surface Optimal Placement in Millimeter-Wave Communications," 2021 15th European Conference on Antennas and Propagation (EuCAP), Dusseldorf, Germany, 2021, pp. 1-5, doi: 10.23919/EuCAP51087.2021.9411076.

\bibitem{IEEEhowto:19} J. Kokkoniemi, J. Lehtomäki and M. Juntti, "Simplified molecular absorption loss model for 275–400 gigahertz frequency band," 12th European Conference on Antennas and Propagation (EuCAP 2018), 2018, pp. 1-5, DOI: 10.1049/cp.2018.0446.

\bibitem{IEEEhowto:20} M. Shehab, B. S. Ciftler, T. Khattab, M. M. Abdallah and D. Trinchero, "Deep Reinforcement Learning Powered IRS-Assisted Downlink NOMA," in IEEE Open Journal of the Communications Society, vol. 3, pp. 729-739, 2022, doi: 10.1109/OJCOMS.2022.3165590.

\bibitem{IEEEhowto:21} 
V. François-Lavet, P. Henderson, R. Islam, M. Bellemare, and J. Pineau (2018), “An Introduction to Deep Reinforcement Learning”, Foundations and Trends in Machine Learning: Vol. 11, No. 3-4, 2018.

\bibitem{IEEEhowto:22}
G E. Uhlenbeck and L S. Ornstein, "On the Theory of the Brownian Motion", Revista Latinoamericana De Microbiologia, 1930.

\bibitem{IEEEhowto:23} M. Elsayed, A. Badawy, A. E. Shafie, A. Mohamed and T. Khattab, "A Deep Reinforcement Learning Framework for Data Compression in Uplink NOMA-SWIPT Systems," in IEEE Internet of Things Journal, vol. 9, no. 14, pp. 11656-11674, 15 July15, 2022, doi: 10.1109/JIOT.2021.3131524.

\bibitem{IEEEhowto:24}
Keller-Gehrig, Walter: Fast algorithms for the characteristic polynomial. Theoretical Computer Science, 36(2-3):309–317, 1985, ISSN 0304-3975. http://dx.doi.org/10.1016/0304-3975(85)90049-0.
x.doi.org/10.1145/345542.345644.

\bibitem{IEEEhowto:25}
V. Vasudevan, M. Ramakrishna, "A Hierarchical Singular Value Decomposition Algorithm for Low Rank Matrices,". 2017-10-08 | Preprint. ARXIV: arXiv:1710.02812v2.

\end{thebibliography}
\end{document}